\documentclass[aps,prx,twocolumn,showpacs,floats,floatfix,superscriptaddress]{revtex4-2}

\usepackage{graphicx}
\usepackage{amsmath}
\usepackage{amssymb}
\usepackage{bm}

\usepackage{color}

\usepackage[english]{babel}
\usepackage[latin9]{inputenc}
\usepackage{graphicx}
\usepackage{multirow}
\usepackage[usenames,dvipsnames]{xcolor}
\usepackage{bm}
\usepackage{mathtools}
\usepackage{rotating}
\definecolor{oceanboatblue}{rgb}{0.0, 0.47, 0.75}
\definecolor{orange}{rgb}{1,0.5,0}
\definecolor{goodgreen}{rgb}{0.1,0.5,0}
\definecolor{goodred}{rgb}{0.7,0,0}
\usepackage[colorlinks,urlcolor=goodgreen,citecolor=blue,linkcolor=goodred]{hyperref}
\usepackage{cleveref}
\usepackage{lineno}
\usepackage{braket}
\usepackage{tikz}
\usepackage{tocvsec2}
\usepackage{scrextend}
\renewcommand\d{\partial}

\newcommand{\sign}{\mathrm{sgn}}

\newcommand{\be}{\begin{equation}}
\newcommand{\ee}{\end{equation}}
\newcommand{\bea}{\begin{eqnarray}}
\newcommand{\eea}{\end{eqnarray}}

\begin{document}

\title{Flux-induced midgap states between strain-engineered flat bands}

\author{Dung X. Nguyen}
\affiliation{Center for Theoretical Physics of Complex Systems, Institute for Basic Science (IBS), Daejeon 34126, Republic of Korea}
\author{Jake Arkinstall}
\affiliation{Department of Physics, Lancaster University, Lancaster, LA1 4YB, United Kingdom}
\author{Henning  Schomerus}
\affiliation{Department of Physics, Lancaster University, Lancaster, LA1 4YB, United Kingdom}
\date{\today}

\begin{abstract}
Half-integer quantized flux vortices appear in honeycomb lattices when the signs of an odd number of couplings around a plaquette are inverted. We show that states trapped at these vortices can be isolated by applying inhomogeneous strain to the system.
A vortex then results in localized mid-gap states lying between the strain-induced pseudo-Landau levels, with $2n+1$ midgap states appearing between the $n$th and the $n+1$st level.
These states are well-defined spectrally isolated and spatially localized excitations that could be realized in electronic and photonic systems based on graphene-like honeycomb lattices.
In the context of Kitaev's honeycomb model of interacting spins, the mechanism improves the localization of non-Abelian anyons in the spin-liquid phase, and reduces their mutual interactions. The described states also serve as a testbed for fundamental physics in the emerging low-energy theory, as the correct energies and degeneracies of the excitations are only replicated if one accounts for the effective hyperbolic geometric induced by the strain. We further illuminate this by considering the  effects of an additional external magnetic field, resulting in a characteristic spatial dependence that directly maps out the inhomogeneous metric of the emerging hyperbolic space. 
\end{abstract}

\maketitle

\section{Introduction}
Topological defects attract attention as they can give rise to robust states that are both spatially and energetically well localized \cite{rev1,rev2,rev3}. The spatial localization arises from the  pinning of the states to surfaces, edges, or points, while the energetic localization often takes the form of exact zero-mode quantization in the middle of a gap, as enforced, \emph{e.g.}, by a particle-hole or chiral symmetry.
In many contexts, the topologically induced states carry unconventional spin, charge, and exchange statistics \cite{anyon0,anyon1,anyon2,anyon3,WenFC1,WenFC2,FCspinliquid,nonabelianvortex,Feiguin2007}.

A prominent example with possible applications in topological quantum computations are the anyons in Kitaev's honeycomb model, an interacting spin system that is exactly integrable when the spin operators are expressed in terms of Majorana fermions \cite{kitaev,MajoranaKitaevExp,reviewMajorana}. In this effective description, the system is equivalent to a homogeneously strained version of graphene that is functionalized by half-integer flux vortices, corresponding to a sign change of an odd number of coupling constants around a plaquette. 
The  homogenous strain opens a gap with anyonic low-energy excitations that can be used to realize the toric code, a paradigmatic platform for topologically protected quantum computation \cite{Nayak08}.
The anyons are bound to the flux plaquettes, and the configurations are topologically protected by time-reversal symmetry, which enforces the coupling constants to remain real. Furthermore, when the gap is closed the system realizes a spin liquid with non-Abelian low-energy excitations that again depend on the flux-vortex configuration \cite{Takagi2019}. 

Here, we combine the topological features of such vortex states with a well-known time-reversal-invariant analogue of a magnetic field, which appears when the effective strain in the system becomes inhomogeneous \cite{guinea}. This pseudomagnetic field leads to the formation of pseudo-Landau levels (pLLs), which have been realized  in experiments on various honeycomb systems \cite{exp1,exp2,exp3,exp4,exp8,Rechtsman2013,exp7,exp5,exp6}.
Notably, the microscopic model dictates an optimal geometry and maximal strain value at which the system becomes exactly solvable in absence of a vortex \cite{cas2014,rachel2016,exp5}.
The pLLs then become exactly degenerate, implying hidden topological and geometric features in a finite, spatially inhomogeneous system.
Identifying the nature and consequences of these features is a central goal of this work. 
In particular, we demonstrate that in the interplay with such inhomogeneous strain, a vortex induces an additional sequence of isolated states that appear in the gaps between the pLLs --- one state between the $0$th and first pLL, three states between the first and the second pLL, and in general $2n+1$ states between the $n$th and the $n+1$st pLL ($n\geq 0$; the sequence is repeated symmetrically for negative energies).

\begin{figure*}[t]
\includegraphics[width=0.8\linewidth]{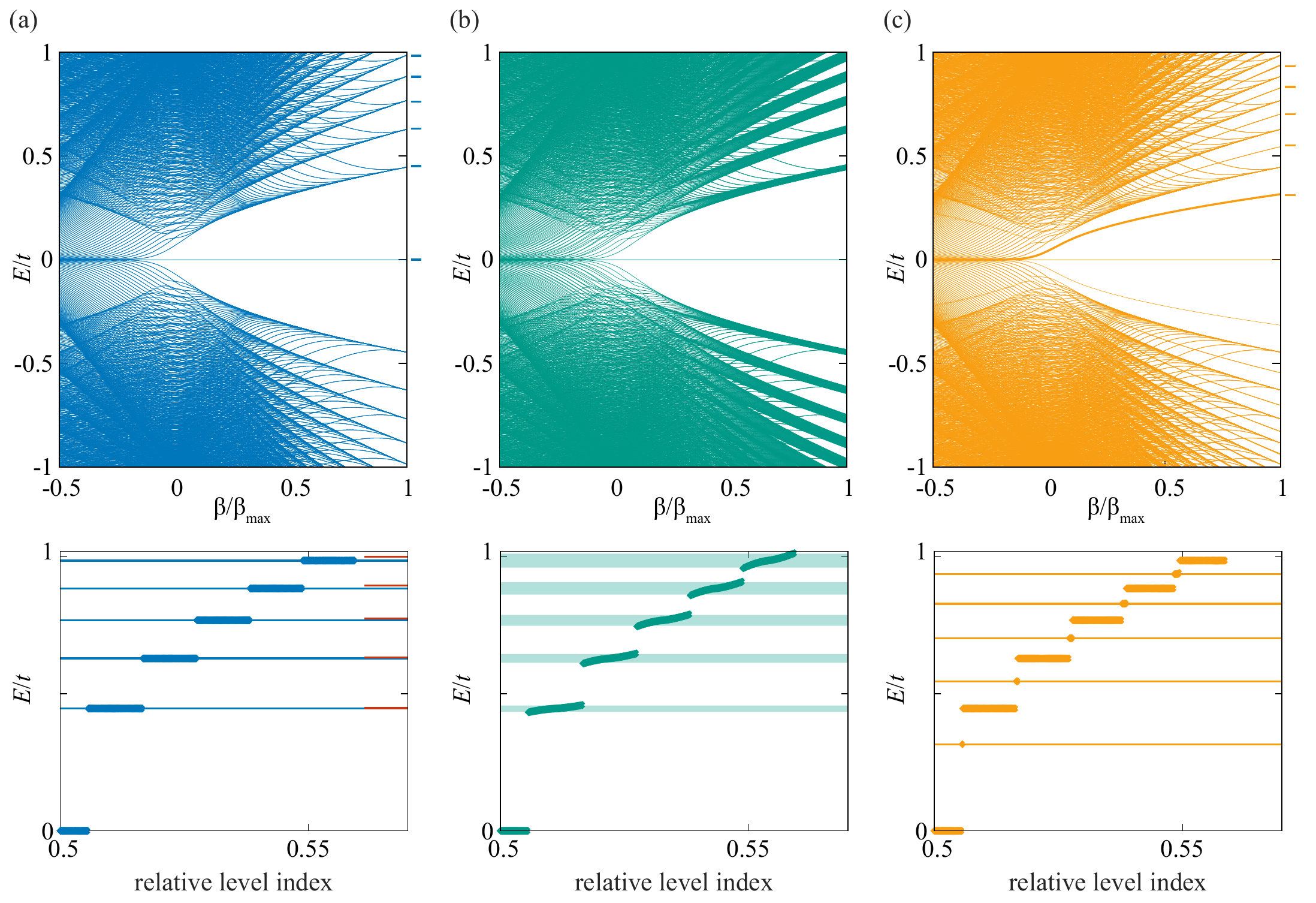}
\caption{\label{fig1}
Strain-dependence of the energy levels of a triangular honeycomb flake (a), in the presence of an additional background magnetic field $B=0.05\beta_\mathrm{max}$ (b), and with a half-integer flux vortex instead placed into the center of the flake (c).
The flake measures 90 hexagons across and is terminated by zigzag edges, conforming to the geometry in Fig.~\ref{fig2}. The strain is given in terms of the strength $\beta$ of the dimensionless pseudomagnetic field (see Eq.~\eqref{eq:straindef}), while the eigenvalues are given in units of the coupling strength $t$ of the pristine system.  As we show in Sec.~\ref{sec:uniform}, the precise energies and degeneracies of the levels in (a) reveal the effects of an emergent curvature in the continuum description of the model. The underlying hyperbolic geometry leads to the broadening of levels in (b), which is in striking contrast to the splitting of the levels expected in conventional low-energy theory, as we describe in Sec.~\ref{sec:uniformmag}. Flux vortices induce a characteristic sequence of midgap states (c), which we describe in Sec.~\ref{sec:flux}, while Sec.~\ref{sec:interactions} contains the application to the Kitaev honeycomb model of interacting spins. 
}
\end{figure*}

We develop these insights by placing the problem into the broader context of the interplay of strain and magnetic fields. This leads us to obtain three key results, which are illustrated in Fig.~\ref{fig1}.
Firstly, we will establish the consistency of the microscopic model with the appropriate continuum theory, which takes the form of a Dirac equation in a curved hyperbolic space \cite{pnuelli1994,WZ,Nguyen2018,Golkar2014}, even though the system is physically flat. The emerging curvature explains the exact position of the pLLs in the optimal strain configuration (see panel(a)), which differs from the conventional estimate
obtained by simply reinterpreting the pseudomagnetic field as a valley-dependent magnetic field  (marked in red). Furthermore, we show that the exact expression for the degeneracy of these levels is consistent with the relativistic Wen-Zee shift \cite{WZ,Golkar2014,Nguyen2018}, which is a topological characteristic of the emerging hyperbolic geometry.
Secondly, we show that the emerging geometry induces a spatial dependence on the effects of an additional external magnetic field, resulting in a broadening of the levels as observed in panel (b). As we will see, this effect directly reflects the inhomogeneous metric of the emerging geometry, which naturally becomes singular at the boundaries of the system.
Thirdly, we show that a flux vortex induces the characteristic sequence of mid-gap states already indicated above, which in panel (c) is illustrated for the case of a half-integer flux vortex placed into the center of the system.
As mentioned, such half-integer flux vortices are of particular interest because they naturally appear  in the Kitaev honeycomb model of interacting spins. Supplementing our findings for such vortices by numerical results for the many-body case, we further establish that the strain also improves the spatial localization of the corresponding many-body excitations, and reduces the range of their mutual interactions.

The paper is organized according to the underlying physical field configurations, which are introduced into the model as described in Sec.~\ref{sec:model}.
In Sec.~\ref{sec:uniform}, we establish the consistency between exact results in the optimally strained tight-binding model and the  continuum theory with an emergent hyperbolic geometry.
In Sec.~\ref{sec:uniformmag}
we describe how this geometry modifies the interplay with an additional external magnetic field.
In Sec.~\ref{sec:flux} we describe the
formation mechanism of the flux-pinned midgap states in the single-particle picture, while Sec.~\ref{sec:interactions} considers the effective interactions of the states and contains the application to the Kitaev honeycomb model of interacting spins.
In the concluding Sec.~\ref{sec:conclusions}, we discuss the general implications of these findings for the interplay of topology and geometry in  finite inhomogeneous systems, and identify further applications. 

Throughout the main text, we focus on key analytical and numerical results that explain and illustrate our findings. 
Appendix \ref{sec:cont} provides background on the continuum theory, Apps.~\ref{sec:unifcont}, \ref{sec:appunifmag}, and App.~\ref{sec:vortcont} applies this to derive results for systems with uniform pseudomagnetic and magnetic fields as well as flux vortices, App.~\ref{sec:pllapp} provides details for the exact construction of pseudo-Landau levels in the microscopic model including a flux vortex, and App.~\ref{sec:addnum} contains additional numerical results.

\begin{figure*}[t]
\includegraphics[width=0.7\linewidth]{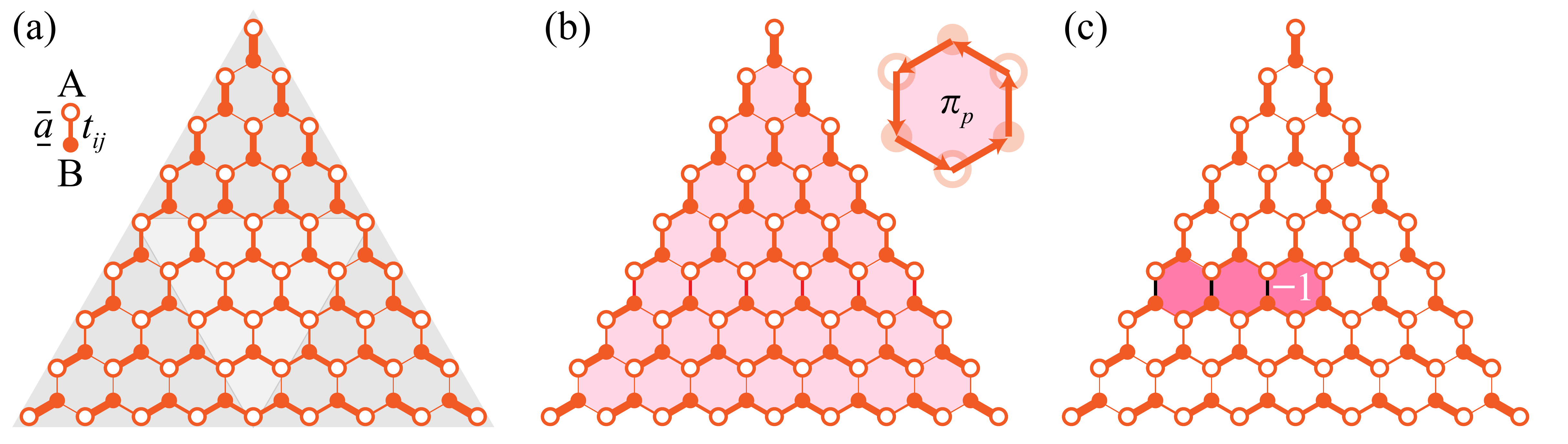}
\caption{\label{fig2}
(a) Zig-zag terminated triangle of a honeycomb lattice in the coupling configuration \eqref{eq:straindef}, which is indicated by the thickness of the lines. This system supports precisely flat pseudo-Landau levels, which serve as the reference for the additional effects from magnetic fields.
We include these magnetic fields via a Peirls substitution either as a uniform background contribution (b), corresponding to fixed plaquette factors $\pi_p$, or as localized vortices (c), including half-integer vortices that naturally appear in the theory of the Kitaev honeycomb model of interacting spins. 
}
\end{figure*}

\section{Model and configurations}
\label{sec:model}
The general theme of this paper is the interplay of magnetic fields and emergent geometry in lattice systems with inhomogeneous coupling configurations.
The microscopic description is provided by a tight-binding Hamiltonian
\begin{equation}
\label{eq:tbham}
H=\sum_{\left<ij\right>}t_{ij} \psi_i^\dagger \psi_j,
\end{equation}
where indices $i$ and $j$ enumerate points on a honeycomb lattice, which we define with bond length $a\equiv 1$ (see Fig.~\ref{fig2}). As indicated there, the lattice is bipartite, allowing to assign sites to two sublattices $A$ and $B$. The field operators $\psi_i$ obey an algebra that depends on whether the underlying system is fermionic, bosonic, or anyonic. As the system is quadratic, the corresponding single-particle picture only depends on the nearest-neighbour couplings $t_{ij}=t_{ji}^*$, which we specify to combine the contributions of effective pseudomagnetic and magnetic fields, including flux vortices at designated positions. 

In the pristine system all couplings are equal, $t_{ij}=t$ with real $t$, resulting in the conventional dispersion relation 
$E(k)=\pm t|1+2e^{i3k_y/2}\cos (\sqrt{3} k_x/2)|$ of graphene \cite{graphenereview}.
In the low-energy theory, this dispersion can then be approximated by Dirac cones $|E|=v_F|\mathbf{k}\mp\mathbf{K}_0|$ around the $K$ and $K'$ points in the Brillouin zone, with $v_F=3t/2$ and $\mathbf{K}_0=(\pm 4\pi/3\sqrt{3},0)$. A precisely uniform pseudomagnetic field of dimensionless strength $\beta$ is obtained by modifying these couplings to \cite{guinea,strain1,juan2012,juan2013,zubkov2015,Henning2013,wagner2022}
\begin{equation}
\label{eq:straindef}
t_{ij}=t\left[1-\frac{\beta}{2}\pmb{\rho}_{ij}\cdot \mathbf{r}_{ij}\right]\equiv t_{ij}^{0},
\end{equation}
where $\pmb{\rho}_{ij}$ is the bond vector from the site $i$ to site $j$ and $\mathbf{r}_{ij}$ is the bond center, both taken in the pristine (unstrained) system.
Notably, in this coupling configuration, commensurate values $\beta=4/N$ with integer $N$ enforce a system of finite size,
as the couplings drop to zero around the edges of a triangle with zigzag edges \cite{cas2014}, leading to the geometry illustrated in Fig.~\ref{fig2}(a).
The microscopic model is then exactly solvable \cite{rachel2016}, displaying a sequence of precisely flat pLLs that serves as the reference point for all results in this paper. 
We discuss this reference configuration in detail in Sec.~\ref{sec:uniform}, where we establish its connection to an emerging hyperbolic geometry.

Our subsequent focus is on the interplay of these strain-induced features with additional magnetic fields, which either act uniformly across the system, as illustrated in  Fig.~\ref{fig2}(b) and discussed in Sec.~\ref{sec:uniformmag}, 
or are localized into flux vortices, as illustrated in  Fig.~\ref{fig2}(c) and discussed in Sec.~\ref{sec:flux}. 
In the microscopic model, these effects are included by a standard Peierls substitution $t_{ij}=t_{ij}^{0}\exp(i\phi_{ij})$  with $\phi_{ij}=-\phi_{ji}$ \cite{Hof1976}. The physical effects are then captured by the flux factors $\pi_p=\prod_p\exp(i\phi_{ij})$, defined by transversing the bonds around the plaquette in a loop with mathematically positive direction as shown in the inset of Fig.~\ref{fig2}(b). 
From all configurations of phases $\phi_{ij}$, only these plaquette fluxes are physically significant when the local $U(1)$ gauge freedom of the fields $\psi_i$ is taken into account. 
A uniform background magnetic field is obtained by setting the fluxes to a common value throughout the system (see Fig.~\ref{fig2}(b)), while localized vortices of arbitrary flux are obtained by modifying the couplings along a line from a plaquette to the edge (see Fig.~\ref{fig2}(c)). In particular, half-integer vortices are modeled by inverting the sign of some of the couplings, $t_{ij}=\sigma_{ij}t_{ij}^{0}$, $\sigma_{ij}=\sigma_{ji}=\pm1$. The plaquette operators $\pi_p=\prod_p\sigma_{ij}$ then take the values $+1$ in the absence of a vortex, and $-1$ in the presence of a vortex. 
By adding the phases $\phi_{ij}$ of such configurations, they can be combined to obtain general vortex patterns above a background magnetic field, while the pseudomagnetic field continues to determine the coupling strengths $|t_{ij}|=t_{ij}^{0}$ according to Eq.~\eqref{eq:straindef}.

\section{Pseudo-Landau levels and hyperbolic geometry}
\label{sec:WZ}
\label{sec:uniform}
We first describe the correspondence of the description of pseudo-Landau levels in the microscopic model and in the appropriate continuum theory. This theory takes the form of a Dirac equation in a curved hyperbolic space \cite{pnuelli1994,WZ,Nguyen2018,Golkar2014},
which we review in Appendix~\ref{sec:cont}. The emerging curvature characteristically affects the energies and degeneracies of the levels, allowing us to recover results from the microscopic model exactly, as we discuss now.

We establish the consistency of both descriptions in the optimally strained reference configuration,
forming a triangular flake terminated by zigzag edges as illustrated in Fig.~\ref{fig2}(a).
As mentioned in the definition of the model, this geometry maximizes the range of the pseudomagnetic field $\beta$ so that the couplings $t_{ij}^{0}/t>0$ remain positive throughout the sample. 
Figure \ref{fig1}(a) shows how pLLs form as a function of the pseudomagnetic field strength in a triangle with $N=90$ sites along each edge.
Within conventional low-energy theory that ignores the curvature effects, the pseudomagnetic length is given by $\ell = 1/\sqrt{|\beta|}$, and the energies are predicted to cluster into pLLs at 
\begin{equation}
\label{eq:enapprox}
\tilde E_n=v_F\,{\rm sgn}\,(n)\,\sqrt{2\beta |n|},
\end{equation}
enumerated by an integer index $n$ with $|n|<N$. 
However, we can improve beyond this simple estimate. 
At the maximal  value $\beta=\beta_\mathrm{max}\equiv 4/N$, the microscopic model can be solved exactly, and the pseudo-Landau levels become precisely flat  \cite{cas2014,rachel2016,exp5} (we recapitulate the construction in App.~\ref{sec:pllapp}).
The energies then take the exact values 
\begin{equation}
\label{eq:energies}
E_n^{(\mathrm{max})}=(2 v_F/N)\,{\rm sgn}\,(n)\,\sqrt{ |n|(2N-|n|)},
\end{equation}
which systematically deviate from $\tilde E_n$ as the level index increases.
Furthermore, the $n$th pLL contains exactly 
\begin{equation}
D_n=N-|n|     
\end{equation}
degenerate levels, displaying a systematic depletion that is absent from conventional Landau levels.

\begin{figure}[t]
    \centering
    \includegraphics[width=\columnwidth]{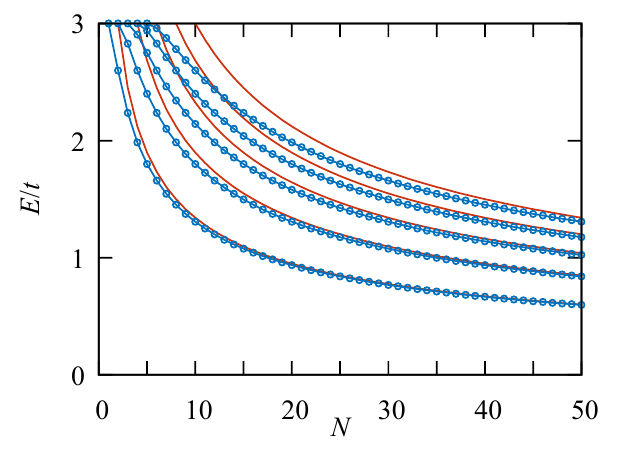}
    \caption{System-size dependence of the pLL energies \eqref{eq:exactmaintext}
at maximal strain $\beta_\mathrm{max}$ in the continuum theory with curvature (thick blue curves), which exactly recovers the result \eqref{eq:energies} of the microscopic model in the optimally strained geometry. The red lines show the conventional estimate \eqref{eq:enapprox} of these levels when the curvature is ignored. For clarity, we only show the levels with index $n=1$ to $5$.}
    \label{fig3new}
\end{figure}

These exact energies and degeneracies match perfectly with the continuum theory. Applied to the coupling configuration \eqref{eq:straindef}, this not only induces a uniform pseudomagnetic field of strength $\mathcal{B}=\beta$, but also a constant negative curvature $ \mathcal{K}=-\beta^2/4$ (see App.~\ref{sec:unifcont}).
The exact energy levels \eqref{eq:energies} then correspond precisely to the Landau levels  
\cite{GGV92,pnuelli1994,wagner2022}
\begin{equation}
\label{eq:exactmaintext}
E_n = v_F \sign(n)\sqrt{2|n \mathcal{B}| + n^2 \mathcal{K}}
\end{equation}
of Dirac fermions on a hyperbolic surface with a constant negative curvature 
\begin{equation}
	 \mathcal{K}=-\frac{\beta_{\mathrm{max}}^2}{4}=-\frac{4}{N^2}
\end{equation}
and a pseudomagnetic field 
\begin{equation}
 \mathcal{B}=\beta_{\mathrm{max}}=\frac{4}{N},
\end{equation}
which are both induced by the space-dependent coupling profile \eqref{eq:straindef}. 
The difference between the exact levels including the curvature and the approximation ignoring the curvature is illustrated in Fig.~\ref{fig3new}.

Furthermore, the degeneracy depletion of the Landau levels exactly matches the relativistic Wen-Zee shift 
\cite{pnuelli1994,WZ,Nguyen2018,Golkar2014}
\begin{equation}
\label{eq:WZ}
	D_n=D_0\left(1+|n| \frac{\mathcal{K}}{\mathcal{B}}\right)=N\left(1-\frac{|n|}{N}\right).
\end{equation}
Here, the degeneracy $D_0=N$ of the 0th pLL can be inferred from the sum rule $\sum_n D_n=N^2$, accounting for all states in Hilbert space.

We note that in the tight-binding model, this degeneracy arises because the specified geometry
contains exactly $N$ more sites on the A sublattice than on the $B$ sublattice. Based on the chiral symmetry of the model, this difference fixes the number of zero modes independent of the coupling configuration.  
Indeed, the curvature does not break the chiral symmetry of the continuum theory, and therefore also preserves the consistency of other symmetry-enforced features, such as the symmetry of the level spectrum itself. Furthermore, the theory recovers that the $0$th pLL is sublattice-polarized on the A sublattice while all other pLLs have equal weight on both sublattices \cite{cas2014,exp5,Henning2013}, which again conforms to the sum rules for these subspaces.  

\begin{figure*}[t]
  \centering
  \includegraphics[width=\linewidth]{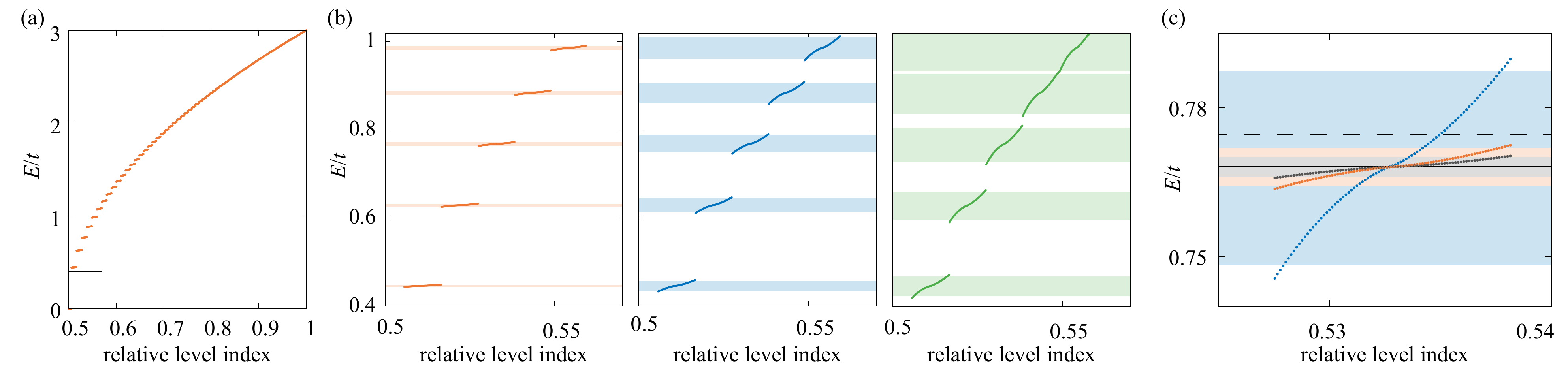}
  \caption{Effect on the pLLs from an additional external magnetic field, demonstrating the curvature-enforced broadening described in the text. (a) The upper half of the numerical level sequence in an optimally strained triangle of size $N=90$, subject to an additional magnetic field of strength $B=0.01\beta_\mathrm{max}$. 
 (b) Comparison of the numerically obtained energies (data points) to the predicted broadening intervals (shaded regions)  for pLLs 1 to 5, at $B/\beta_\mathrm{max}=0.01, 0.05, 0.1$ (subpanels, left to right). (c) Detailed broadening of the 3rd pLL, for $B/\beta_\mathrm{max}=0.05, 0.01, 0.05$ (gray, orange, and blue). The solid black line shows the exact position \eqref{eq:exactmaintext} of the pLL at $B=0$, while the dashed line is the conventional approximation \eqref{eq:enapprox} ignoring the curvature.
  }\label{fig4}
\end{figure*}

These observations are remarkable because we derived the continuum theory by expanding around the center of the system, where we can treat both low-energy valleys separately.
The theory then approximates the system as a hyperbolic disk of radius $r = 4/|\beta|$, defined by the distance where the metric becomes singular, which happens to coincide with the finite system size of the optimally strained microscopic system.
Choosing a different expansion point, we can extend these considerations qualitatively to capture additional  features, such as  
the opening of a local gap in the regions around the corner of the triangle. This then explains the spatial support of the low-lying pLLs, which fill out an approximately triangular region in the center of the system  \cite{cas2014,exp5} (see  Fig.~\ref{fig11} in the numerical Appendix \ref{sec:addnum}).

Overall, the continuum theory therefore captures global and local features of the pLL spectrum and states, and reveals their connection to geometric curvature effects. In turn, the underlying microscopic model realizes these often elusive effects precisely.
This provides the platform on which we can now include the effects of additional magnetic fields, extended across the system or localized in vortices.

\section{Interplay with a uniform external magnetic field}
\label{sec:uniformmag}

As we establish next, the strain-induced hyperbolic geometric affects the interplay of the pseudomagnetic field with an additional external magnetic field $B$, even when the latter is physically uniform across the system.
In general, both types of fields can be distinguished by their symmetry properties. A real magnetic field breaks time-reversal and parity symmetry, and enters with the same sign in the two sectors of the continuum theory, which are associated with the low-energy valleys near the K and K$'$ points. Furthermore, as in these sectors, the role of the sublattices is interchanged, the 0th Landau level from a magnetic field is not sublattice polarized. In contrast, the pseudomagnetic field preserves both symmetries, but switches its sign between both valleys, and therefore supports the sublattice-polarized 0th level, as discussed above. 
The conventional low-energy theory would therefore predict that an additional magnetic field splits each pLL into two distinct levels.
However, as we will discuss now, the geometric curvature changes this picture drastically, so that one instead obtains continuously broadened pLLs that remain centered at the value without a magnetic field, as already illustrated in Fig.~\ref{fig1}(b). 

As derived in Appendix \ref{sec:appunifmag}, this additional feature becomes visible when one combines the two fields in the continuum theory. 
We find that the curvature remains intact at $\mathcal{K}=-\beta^2/4$. 
However, the combined effective field 
\begin{equation}
	\mathcal{B}^{K}=\beta+\tilde{B}, \quad \mathcal{B}^{K'}=-\beta+\tilde{B}
\end{equation} 
seen by Dirac fermions near $K$ and $K'$ points features an effectively space-dependent  magnetic field
\begin{equation}
    \tilde{B}=\frac{B}{\sqrt{g}}= B |1-\beta^2(x^2+y^2)/16 |.
\end{equation} 
As indicated, this spatial dependence directly tracks back to the inhomogeneous metric of the system, which is encoded in the quantity $g$.
We note that this expression directly encodes the distance $r = 4/|\beta|$ at which the metric in the continuum theory becomes singular. 
The effective magnetic field $\tilde{B}$ then varies from $B$ near the center to $0$ near the boundary of the system.

At the center of the sample, the combined effective field therefore takes the valley-dependent magnitude $|\mathcal{B}|=|\pm\beta + B|$, but at the boundaries it reduces to the valley-independent magnitude $|\mathcal{B}|=|\beta|$ from the pseudomagnetic field.
As the effective magnetic field varies slowly over the sample and the underlying flat-band states can be defined with a local support, we obtain the prediction that instead of becoming split, the energies become spread out between two values 
\begin{align}
\label{eq:EBK}
	E_n^K &=v_F {\rm sgn}\,(n)\sqrt{2|n(\beta+B)|-n^2\frac{\beta^2}{4}}, \\
\label{eq:EBKp}
	E_n^{K'}&=v_F {\rm sgn}\,(n)\sqrt{2|n(-\beta+B)|-n^2\frac{\beta^2}{4}}, 
\end{align}
where we again included the contributions from the curvature. This prediction is expected to be valid as long as the perturbed Landau levels remain well separated, which implies that $|n|\lesssim |\beta/B|$, and hence also that the magnetic field is not too strong.
Higher up in the spectrum, we encounter a continuous spectrum, which can be physically attributed to magnetic edge states filling out all gaps (see Fig.~\ref{fig10} in the numerical Appendix for further illustration).

This picture is confirmed in Fig.~\ref{fig4}, where we compare the predicted broadening to numerical results for optimally strained triangles with an additional magnetic field of various strengths. This broadens the levels as expected, with an inflection point at the energy \eqref{eq:exactmaintext} of the system without the additional field. As already mentioned, this behavior is in striking contrast to the conventional low-energy prediction without curvature, which would result in the splitting of each pLL into two distinct levels.
A well-defined splitting can only be observed when the system is studied locally, as done experimentally, e.g., in Ref.~\cite{exp4}, and this observation also prepares our study of flux vortices in Secs.~\ref{sec:flux} and \ref{sec:interactions}.

\section{Flux vortices}
\label{sec:flux}

As already mentioned in the introduction,
the interplay of strain and magnetic fields takes a particularly striking form when the latter are localized in flux vortices. 
As illustrated in  Fig.~\ref{fig1}(c) for a  half-integer flux vortex placed into the center of the system, this induces a characteristic sequence of midgap states, where we find 1 state between the $0$th and first pLL, 3 states between the first and the second pLL, and in general $2n+1$ states between the $n$th and the $n+1$st pLL. 
This goes along with a reduction of the degeneracy in the pLLs by $2|n|$ (therefore, the $n$th pLL contains  $ N-3|n|$ states). This indicates that from each pLL, $|n|$ states each are donated to the  gap above and below. For example, the seven midgap states between pLLs 3 and 4 then correspond to the combination of $3$ states missing from the $3$rd pLL and  $4$ states missing from the $4$th pLL. 
In the following, we will confirm this picture in detail. For this, we first explain the modified degeneracy pattern of the pLLs microscopically in the optimal strain configuration. Then, we utilize the continuum theory to describe the formation of the midgap states themselves.

\subsection{Flux-modified pLL degeneracy}
\label{sec:fluxa}

To explain the modified degeneracy pattern of the pLLs in the presence of a flux vortex, we adapt the construction principle of these states in the case without a flux \cite{rachel2016}. 
As further shown in \cite{exp5}, the original construction can be used first to determine a zero mode localized on the A sites along one of the edges of the triangle, and then construct the remaining zero modes recursively by including A sites on successive parallel lines. With Ref.~\cite{rachel2016}, the states in the finite-energy pLLs are then obtained by combining such solutions recursively from triangles of different sizes, where one level is lost in the passage from one pLL to the next. 

\begin{figure}
    \centering
    \includegraphics[width=\columnwidth]{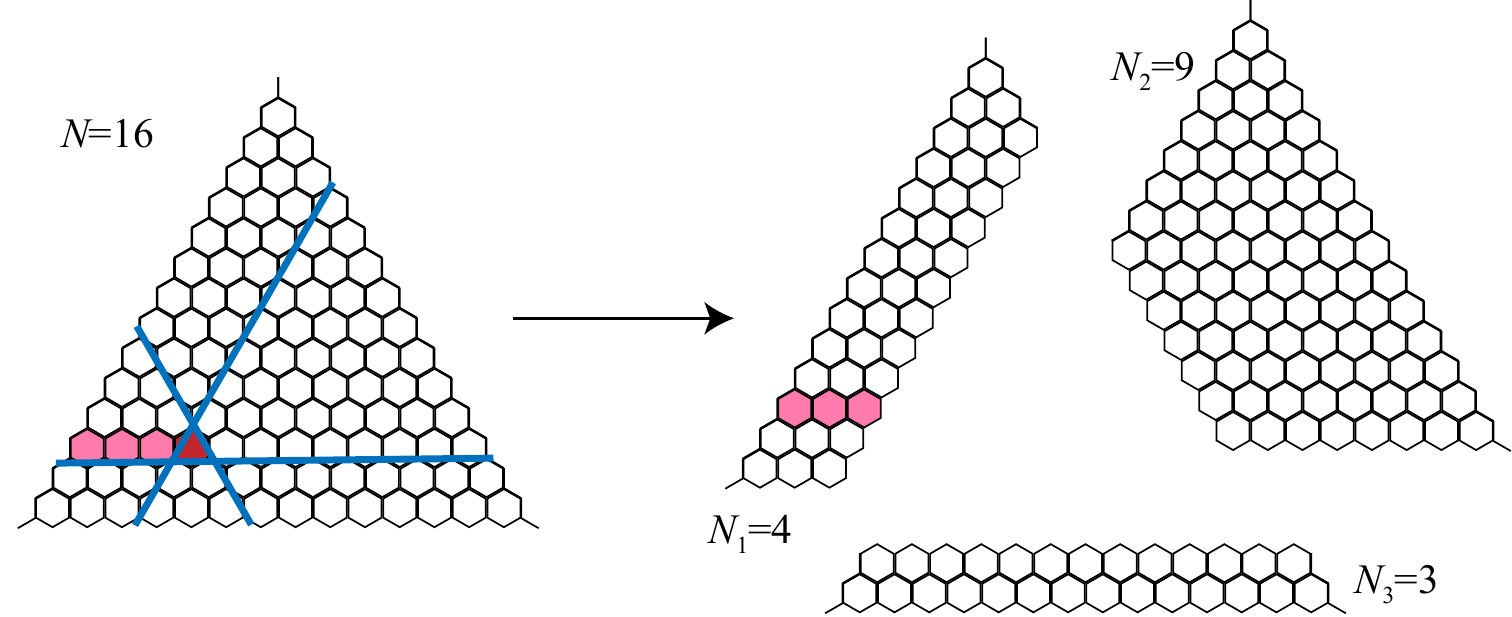}
    \caption{Construction principle of exact pLL states in an optimally strained triangle with a flux vortex, marked in red, which is obtained by a Peirls substitution along the line of the shaded plaquettes. The desired states are obtained by adopting their construction principle in absence of the vortex, which can be used to produce a basis of zero modes with trapezoidal support. We use this to construct three trapezoidal systems whose pLLs are exact solutions of the triangle with the vortex, but are not affected by its existence (the remaining shaded plackets do not carry any flux, so this also holds true for subsystem 1).  The combined Wen-Zee shift from the three systems exactly accounts for the observed number of midgap states produced by the vortex.}
    \label{fig5}
\end{figure}

We recapitulate and amend this construction in detail in Appendix \ref{sec:pllapp}.
In the following, we only need to utilize that this can be used to generate exact pLL states with compact trapezoidal support. 
This implies that the construction partially carries over to a system with a flux vortex, as long as one avoids crossing the corresponding plaquette. We therefore proceed as illustrated in Fig.~\ref{fig5}, and approach the vortex from three sides. This produces three  separate sets of $N_i$ zero modes that sum to $\sum_i N_i=N$. As indicated in the figure, each set corresponds to the 0th pLL of a  system of reduced size, with the shape modified into a trapezoid. 
The string of modified couplings connecting the flux plaquette to the boundary only affects one  subsystem, and crosses it completely, so that it can be gauged way. For each of these subsystems, we can then construct $N_i-|n|$ states in the higher pLLs recursively as in the original approach, where again one level is lost in the
passage from one pLL to the next. These particular states then provide solutions of the triangular system including the vortex, for arbitrary flux (the subsystems also contain additional states, as their combined Hilbert space dimension is larger than in the original system).
Therefore, the degeneracy of each pLL in the triangular system with the vortex obtained in this way is $\sum_i N_i-|n|=N-3|n|$.
This recovers the observed reduced degeneracy, with the deficit of states given by the midgap states.

As we will show next, these midgap states can be exactly accounted for in the continuum theory. Specifically, from each pLL in the vortex-free system, $|n|$ such states are donated to the upper gap, while an equal number of $|n|$ states are donated to the lower gap. Therefore, in the continuum theory, the same modified degeneracy of pLLs is obtained by a remarkable combination of the Wen-Zee shift with the number of vortex-induced midgap states.

\begin{figure}[t]
  \centering
  \includegraphics[width=\columnwidth]{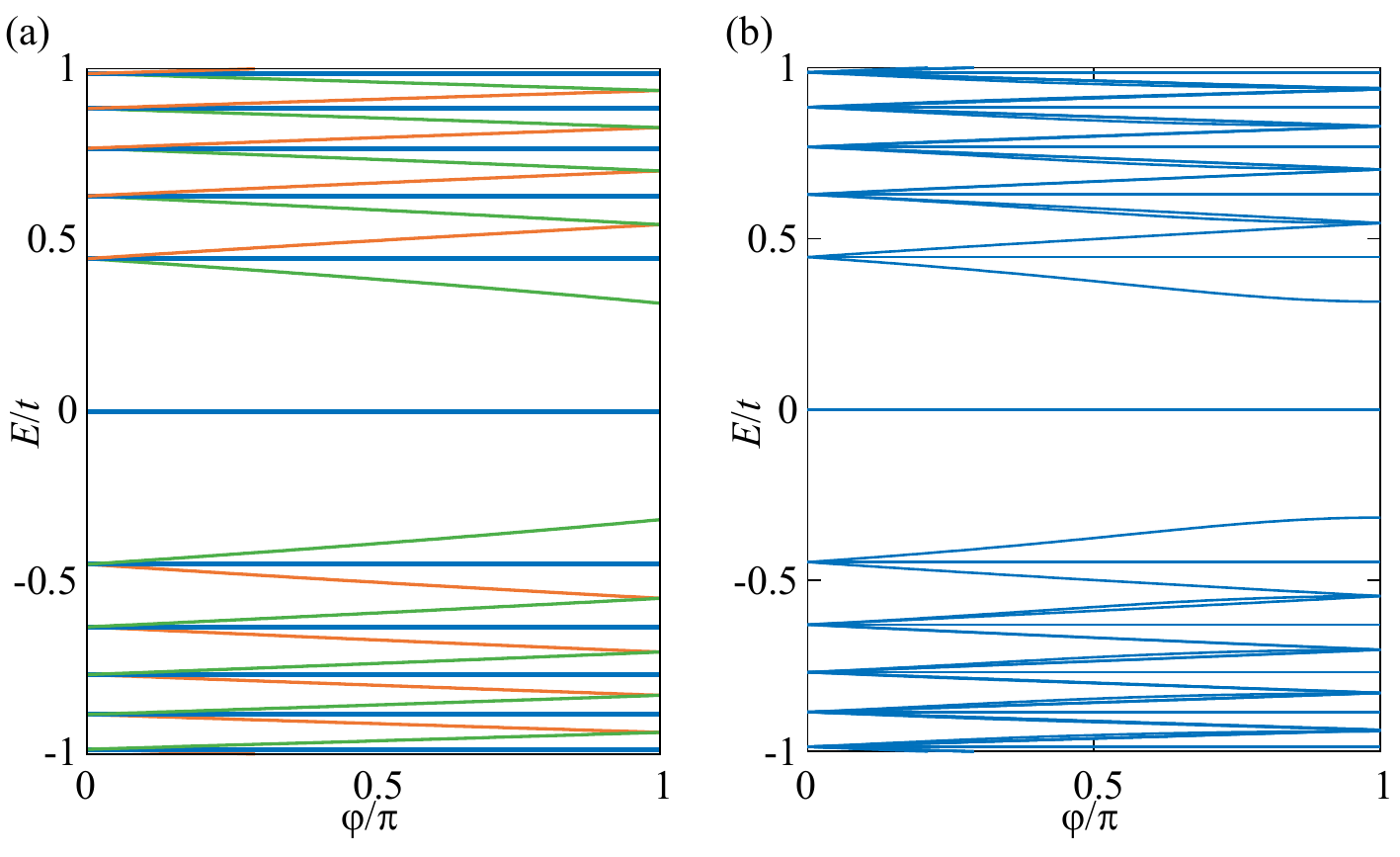}
  \caption{Formation of flux-induced midgap states
  in an optimally strained triangle ($N=90$),
  as a function of the flux $\varphi$ in a vortex placed at its center. Panel (a) shows the analytically predicted midgap energies from Eqs.~\eqref{eq:EvKm} (orange) and \eqref{eq:EvKpm} (green) along with the pLL energies Eq.~\eqref{eq:exactmaintext} (blue), while panel (b) shows the numerical result from the microscopic model.}\label{fig6}
\end{figure}

\subsection{Vortex state formation}
We now turn our attention to the midgap states themselves. For this, we utilize exact solutions in the continuum theory, for a flux of arbitrary value $\varphi$ that is placed into the center of the system. The detailed derivations are given in Appendix  \ref{sec:vortcont}. 
For each valley, we obtain a sequence of midgap states with energies following from Eqs.~\eqref{eq:EvKcur}, \eqref{eq:EvKpcur}, giving
\begin{align}
\label{eq:EvKm}
	E_{\mathrm{gap},n}^K&= v_F \mathrm{sgn}\,(n) \sqrt{2|(|n|+\frac{\varphi}{2\pi})\beta|-(|n|+\frac{\varphi}{2\pi})^2\frac{\beta^2}{4}}, \\
\label{eq:EvKpm}
	E_{\mathrm{gap},n}^{K'}&= v_F \mathrm{sgn}\,(n) \sqrt{2|(|n|-\frac{\varphi}{2\pi})\beta|-(|n|-\frac{\varphi}{2\pi})^2\frac{\beta^2}{4}}. 
\end{align}
These expressions reflect that the flux seen in the two  valleys is the same, while the pseudomagnetic field strength takes opposite values. 
For $\varphi\to 0$, the energies reduce to the positions of the pLLs
\eqref{eq:exactmaintext}, so that the index $n$ traces the origin of these states. 
The flux then shifts the quantization condition for the midgap states, consistent with the extra phase picked up on a cyclotron-like orbit around the vortex, which is fully born out by the analytical form of the derived wavefunctions. 
Furthermore, each of these energies describes $|n|$ midgap states, which in the continuum theory are exactly degenerate.
Within each gap, we therefore find $|n|$ midgap states arising from one valley, and $|n|+1$ states arising from the other valley, which accounts for all the expected states. 

These predictions are confirmed in 
Fig.~\ref{fig6}, where we compare Eqs.~\eqref{eq:EvKm} and \eqref{eq:EvKpm} with numerical results for an optimally strained triangle for size $N=90$, as a function of the flux $\varphi$ in a vortex placed at the center of the system.
We note that each valley produces a symmetric spectrum, in keeping with the chiral symmetry of the system. 

We also see that the midgap states from different valleys become degenerate at half-integer flux, $\varphi=\pi$. According to Eqs.~\eqref{eq:EvKm} and \eqref{eq:EvKpm}, within each gap the energies of states donated from both valleys then indeed coincide,
$E_{\mathrm{gap},n}^K=E_{\mathrm{gap},n+1}^{K'}$,
giving
\begin{equation}
\label{eq:engv}
E_{\mathrm{gap},n}^{(\mathrm{max})}=
(2 v_F/N)\,{\rm sgn}\,(n)\,\sqrt{ (|n|+1/2)(2N-|n|-1/2)}
\end{equation}
at maximal strain.
These levels are indicated by the lines in the bottom panel of Fig.~\ref{fig1}(c).

As we mentioned in the introduction, such vortices appear naturally in the Kitaev honeycomb model of interaction spins, leading to localized excitations with unconventional particle statistics.  We therefore now turn our attention to the effects of the strain on these interactions. 

\begin{figure}[t]
\includegraphics[width=\columnwidth]{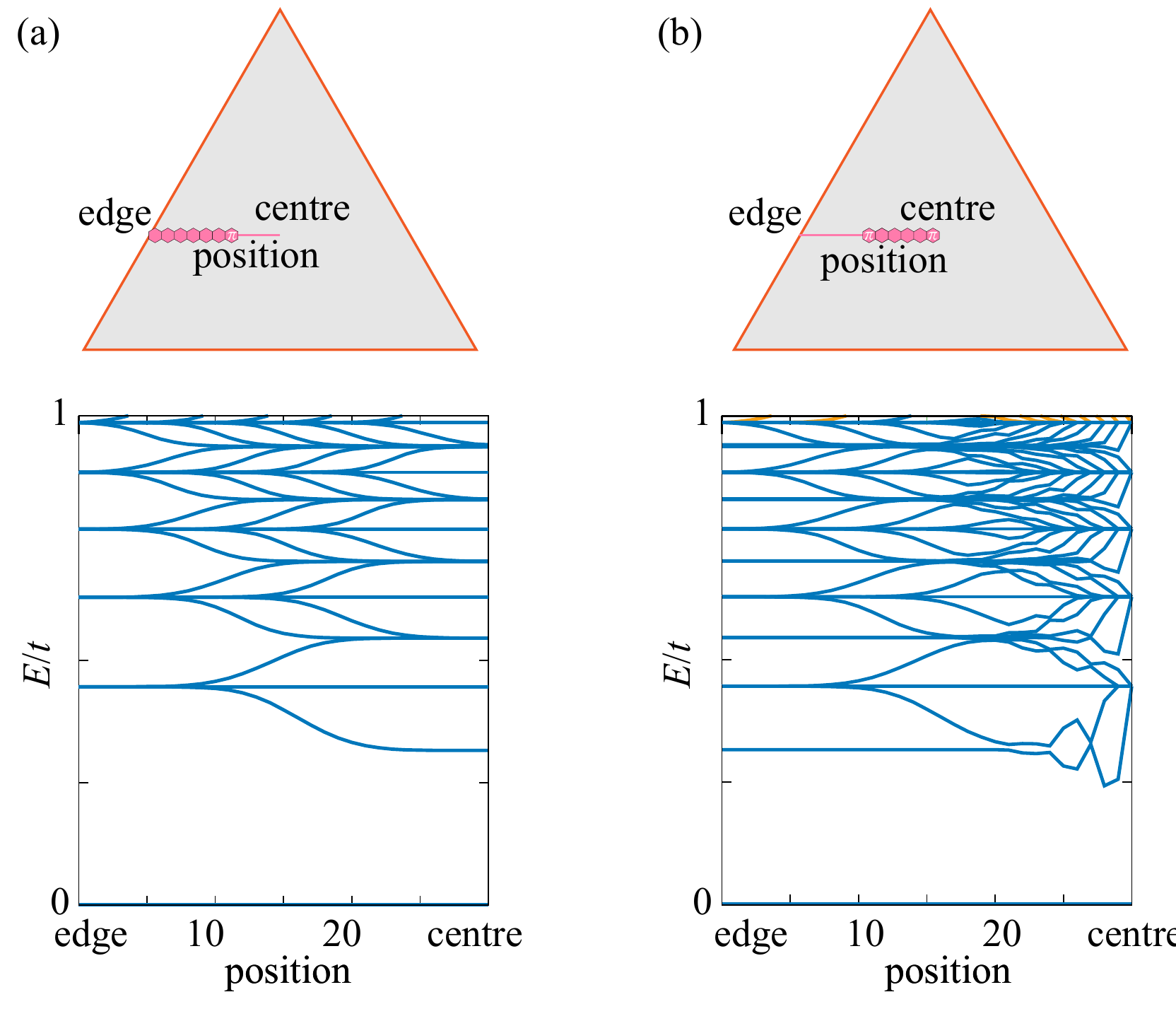}
\caption{\label{fig7}
Single-particle energy levels in an optimally  strained triangle of size $N=90$, as a function of vortex positions. In (a), a single half-integer vortex is placed at a position along a line from the edge to the center. In (b), the system contains an additional half-integer vortex fixed at the center.
}
\end{figure}

\section{Vortex spectroscopy and the Kitaev honeycomb model}
\label{sec:interactions}

We now combine the insights from the previous sections and investigate the position-dependent interplay of flux vortices, first in the single-particle picture, and then in the application to the Kitaev honeycomb of interacting spin \cite{Singh2012}. 
Guided by our discussions in Secs.~\ref{sec:WZ}, \ref{sec:uniformmag}, \ref{sec:flux}, we expect that the flux vortices allow us to probe the local physical features of the system, such as the opening of a local gap in the triangle corners, and the inhomogeneous emerging metric that reduces the effects of a magnetic field near the boundaries. Our particular attention is on  the localization of the vortex states themselves, as this is directly influenced by the inhomogeneous strain, and induces the most characteristic energetic signatures.
 
\subsection{Single-particle picture}

We first establish these effects in the single-particle picture.
The explicit form of the wavefunctions in the continuum theory demonstrates that the pseudomagnetic field localizes the midgap states at the flux vortex, where the localization length is given by the pseudomagnetic length  $\ell=1/\sqrt{|\beta|}$.
Each cluster contains a maximally localized state, while additional states are reminiscent of excited states within an effective potential well.
This general picture is confirmed by our supplementary numerical results,
depicted in Fig.~\ref{fig12}. 

This observation suggests to probe the spatial localization of these states spectroscopically by placing the vortices at different positions.
Figure \ref{fig7}(a) depicts the energetics  when a vortex is placed along a line from the edge to the origin. The midgap states appear in well-defined transitions, denoting the positions where they move fully into the system. In turn, we can interpret this effect as a lifting of their degeneracy as the states are moved toward the edge of the system. This degeneracy lifting, therefore, reveals the differing finite extent of the states, which is consistent with the described sequence of states within each cluster. 

In panel (b) we show the energetics for two vortices, one placed into the center and another moved in from the side. This reveals that each of them pins states separately as long as their extent does not significantly overlap. Such multi-vortex configurations appear naturally in the Kitaev honeycomb model \cite{vortexKitaev1,vortexKitaev2,vortexKitaev3}, to which we turn next.

\begin{figure}[t]
\includegraphics[width=\columnwidth]{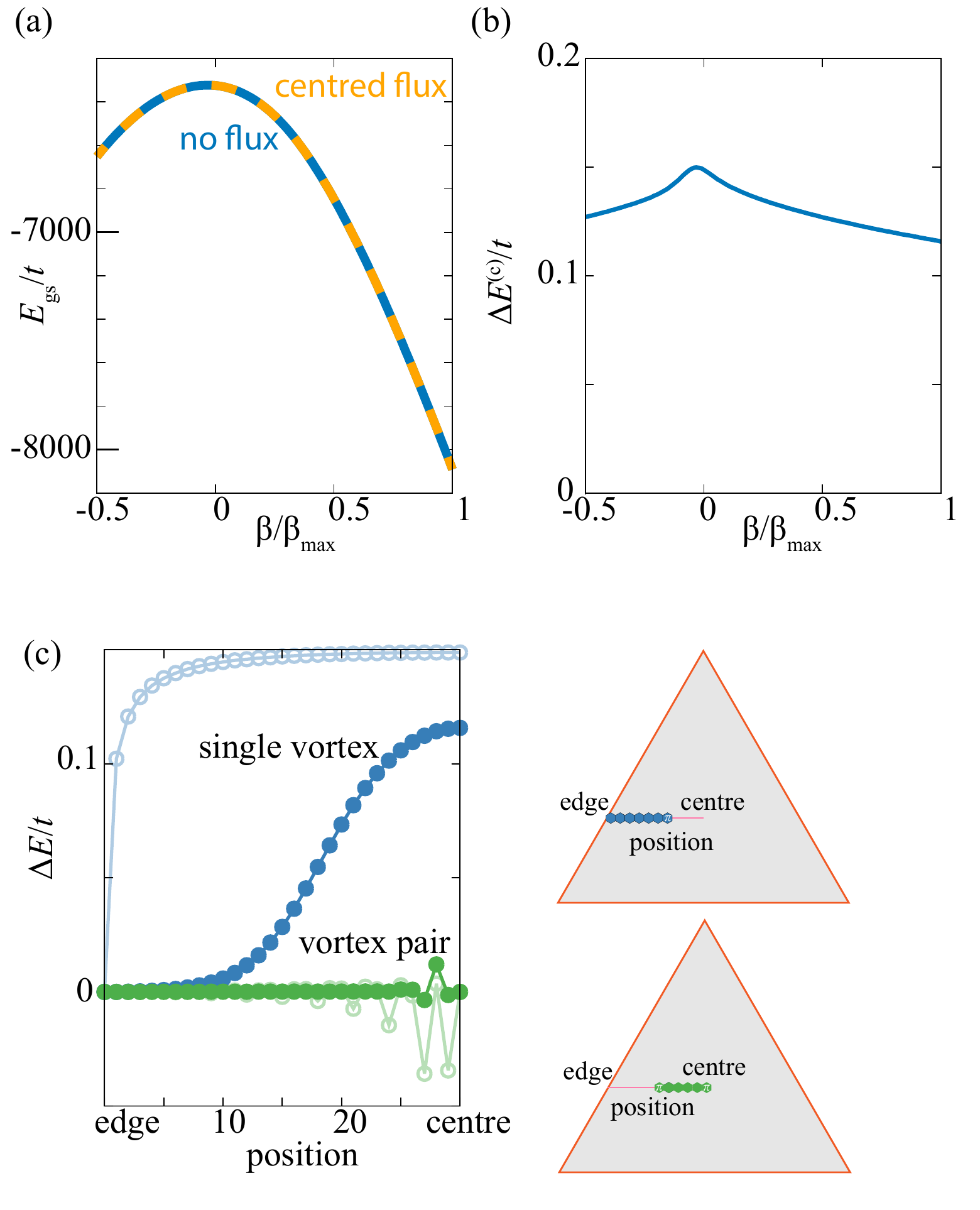}
\caption{\label{fig8}
Vortex energetics in the Kitaev honeycomb model. (a) Strain dependence of the ground-state energies in the flux-free sector and in the sector with a flux vortex at the center of the system. (b) Excess energy from the centered vortex.
(c) Position dependence of excess energies due to a single vortex or a vortex pair in the optimally strained system (full symbols) and in the unstrained system (faint symbols). The results for the vortex pair represent an effective interaction energy, as defined in Eq.~\eqref{eq:vint}.}
\end{figure}

\subsection{Kitaev honeycomb}
\label{sec:kitaev}
In the Kitaev honeycomb model, the tight-binding Hamiltonian \eqref{eq:tbham} arises after a transformation to  Majorana fermions \cite{kitaev,Chen:2008}. For each fixed configuration of vortices (sector), the single-particle energies occur in pairs $\pm\varepsilon_n$, and the  many-body energies $E_{\{\sigma_n\}}=\sum_n \sigma_n \varepsilon_n$ of the original spin states are obtained by summing up these energies for all combinations of signs $\sigma_n=\pm 1$.
The overall ground state is obtained from the vortex-free sector. For homogenous uniaxial strain with a coupling strength $t_3>t_1+t_2$ along one bond direction that exceeds the combined couplings along the other two directions, a gap opens in the vortex-free sector, which can be populated by zero modes in the sectors with vortices; these excitations are then found to obey anyonic statistics. In the inhomogeneously strained setting considered here, the vortex-free sector is not gapped, since the 0th pLL contributes a highly degenerate set of states around $\varepsilon_n=0$, corresponding to a spin liquid \cite{Takagi2019}. Thus, the system possesses a large number of low-energy excitations, which are part of a quantized excitation spectrum.
This quantized excitation spectrum is then modified by the vortex-induced states residing between the pLLs.

These energetic features are illustrated in  Fig.~\ref{fig8}. 
As shown in panel (a) the overall ground state energy $E^{(0)}_{\mathrm{gs}}(\beta)$ from the vortex-free sector systematically depends on the pseudomagnetic field strength $\beta$, displaying a maximum at $\beta=0$. Furthermore, a very similar dependence is shown by the lowest-energy state $E^{(\mathrm{c})}_{\mathrm{gs}}(\beta)$ in the sector with a single vortex, placed into the center (`c') of the system. 
The energy difference 
\begin{equation}
    \Delta E^{(\mathrm{c})}(\beta) = E^{(\mathrm{c})}_{\mathrm{gs}}(\beta)
    -
    E^{(0)}_{\mathrm{gs}}(\beta),
\end{equation}
shown in panel (b), is much smaller than the overall scales for these energies and their strain dependence. Furthermore, as required, it is always positive. 

In Fig.~\ref{fig8} (c) we further analyze the excess energies 
$\Delta E^{(n)}(\beta_{\mathrm{max}})$ at maximal strain for a single vortex that is placed at 
different positions $n$, along the line indicated in the illustration. We observe that the excess energy becomes small when the vortex is moved from the center to the edge. The vortex then resides in a region where the system is locally gapped. 
This behaviour is in contrast with the unstrained system, for which the excess energy drops off only very close to the edge (faint open symbols).

Figure \ref{fig8}(c)  also examines the energetics for a system with two vortices, where one is placed into the center and the other moved along the specified line. When the two vortices are well separated, we find that the excess energy is well approximated by the sum of the individual excess energies (see the supplementary numerical results in Fig.~\ref{fig13}). 
We therefore define the residual energy 
\begin{equation}
\label{eq:vint}
    \Delta E^{(c,n)}=E^{(c,n)}_{gs}-E^{(c)}_{gs}-E^{(n)}_{gs}+E^{(0)}_{gs},
\end{equation}
which can be interpreted as an effective interaction energy.
As shown in the figure, this residual energy is very small, and short-ranged, in particular when compared to the unstrained system (again indicated by faint open symbols).

\section{Conclusions and outlook}
\label{sec:conclusions}

In summary, we have studied the interplay of inhomogeneous strain and magnetic fields in honeycomb models.  This reveals a number of striking effects.
In the appropriate continuum theory, the strain not only induces a valley-dependent pseudomagnetic field, but also an effective negative curvature. As we established in Sec.~\ref{sec:uniform}, this curvature directly affects the energies and degeneracies of the strain-induced pseudo-Landau levels. The emerging  hyperbolic geometry also renormalizes the effects from an additional magnetic field, which becomes weakened at the natural boundaries of the system, as we demonstrated in Sec.~\ref{sec:appunifmag}. The interplay of effects  takes a particularly striking form when the magnetic field is localized into flux vortices. As shown in Sec.~\ref{sec:flux}, this generates a characteristic sequence of midgap states, which reflects further topological and geometric features of the system.
As an application, we investigated in Sec.~\ref{sec:interactions} how these states can be used spectroscopically to probe the local features of the system. In the Kitaev honeycomb system of interacting spins,  the energetics of the vortex states map out the phase diagram in space, and the pseudomagnetic field leads them to exhibit an enhanced spatial localization and short-ranged pairwise interactions in the  spin-liquid phase.

Honeycomb lattice models can also be physically realized for fermions in graphene, where large pseudomagnetic fields can be achieved by in-plane strain \cite{exp1,exp2,exp3,exp4}, and a range of analogous bosonic systems including photons in optical waveguide lattices and resonator arrays, where strain-induced pLLs where realized in Ref.~\cite{Rechtsman2013,exp5,exp6}, while half-integer flux vortices where engineered in Ref.~\cite{Keil16}. 
Our results therefore directly apply to the strain engineering in these settings. For instance, we predict that the splitting of Landau levels from strain and magnetic fields varies spatially, which explains why this effect remained elusive until systems were probed locally as in Ref.~\cite{exp4}. To create effective half-integer flux vortices in graphene, we propose to consider chemical functionalization via adatoms that bridge over bonds, which replicates the photonic mechanism in Ref.~\cite{Keil16}.

For further investigations, we therefore suggest seeking the physical signatures of the strain-localized flux-induced midgap states in these three settings. In photonic systems, these states could be used for waveguiding.  In electronic systems,  they can trap charges and serve as scattering centers, which can be investigated in transport. In the context of the Kitaev honeycomb model, it would furthermore be particularly interesting to establish their implications for the dynamics in time-dependent systems, including for braiding. 
Further methodological progress could target the modification of the continuum theory to include the interplay of the valleys, and ultimately also the collision of the Dirac points and the opening of a gap. Our microscopic considerations furthermore suggest significant scope to generalize exactly solvable microscopic models, both in terms of their geometric shape as well as in terms of their physical properties.

\begin{acknowledgments}
HS thanks the Center for Theoretical Physics of Complex Systems at the Institute for Basic Science in Daejeon for the kind hospitality during the main phase of this research.  D.X.N. is supported by Grant No. IBS-R024-D1.
\end{acknowledgments}

\appendix

\section{Emergence of curved geometry in the continuum theory}
\label{sec:cont}
In this Appendix we provide background information on the continuum theory employed throughout this work, detailing in particular how the strain does not only induce a pseudomagnetic field, but also leads to an effectively curved geometry. For this background material, we follow Refs.~\cite{wagner2022,juan2013}.
\subsection{Notation}
We first fix the notation. We define the spatial metric as 
\begin{equation}
ds^2=g_{ij}dx^i dx^j
\end{equation}
and fix the flat-space metric as 
\begin{equation}
\eta_{ij}=\delta_{ij}.
\end{equation}
We will use $a,b,c,\cdots$ for local frame indices and $i,j,k, \cdots$ for space coordinate indices. The vielbein is given by the definition 
\begin{equation}
g_{ij}=e^a_i e^b_j \eta_{ab},
\end{equation}
which corresponds to a transformation to a locally flat frame. 
We lower and raise the local frame indices by $\eta_{ab}$ and $\eta^{ab}$, and lower and raise the space coordinate indices by $g_{ij}$ and $g^{ij}$. Furthermore, we introduce the inverse vielbein via the definition $e^a_j e^j_b=\delta^a_b$.

With this data, the spin connection is defined as 
\begin{equation}
\omega_i =\frac{1}{2}\epsilon_{ab}e^{aj}\nabla_i e^b_j,
\end{equation}
where the action 
\begin{equation}
\nabla_i e^a_j=\partial_i e^a_j -\Gamma^k_{ij}e^a_k
\end{equation}
of the covariant derivative on a 1-form vielbein involves the Christoffel symbol
\begin{equation}
\Gamma^i_{kl}=\frac{1}{2}g^{im}\left(\frac{\d g_{mk}}{\d x^l}+\frac{\d g_{ml}}{\d x^k}-\frac{\d g_{kl}}{\d x^m}\right)=\frac{\d e_k^a}{\d x^j}e^i_a
.
\end{equation}
 We also use the explicit notation 
\begin{equation}
\gamma^0=\sigma^3, \qquad \gamma^i=\sigma^3 \sigma^i \qquad (i=1,2)
\end{equation}
of gamma matrices in 2+1 dimension,
where $\sigma^i$ denotes the standard Pauli matrices. 

\subsection{Strain-induced pseudomagnetic field and curvature}

We now bring the tight-binding model \eqref{eq:tbham} into a form where it can be interpreted as a Dirac equation with an induced pseudomagnetic field in curved space.
As a first step, we write the general coupling profile 
in the form 
\begin{equation}
t_{ij} \approx t e^{i A_{ij}}\left[ 1 - \tilde{\beta} (\mathbf{u}_j-\mathbf{u}_i)\cdot\pmb{\rho}_{ij} \right]\label{hop},
\end{equation}
which allows us to interpret its spatial modulation physically in terms of a displacement field $\mathbf{u}_i$ with  effective coupling strength 
$\tilde{\beta}$, and connects the phases $A_{ij}=\pmb{A}\cdot \pmb{\rho}_{ij}$ to an external vector potential with components $A_i$. In these expressions, $\pmb{\rho}_{ij}$ continues to refer to the pristine configuration. 
To first order in the strain,
the effective low-energy Hamiltonian near the $K$ point is then given by \cite{wagner2022}
\begin{equation}\label{HamTB}
\mathcal{H}_{TB} = v_F \left[ i  v_{ij}(\mathbf{r}) \sigma_i (\partial_j-iA_j) + i \sigma_i \Gamma_i  + \sigma_i
A^s_i \right],
\end{equation}
where $v_{ij}$ is the space-dependent modulation of the  Fermi velocity $v_F=\frac{3t}{2}$, 
\begin{equation}
\Gamma_i = \frac{1}{2}  \d_k v_{ik},
\label{Aprima}
\end{equation}
and $A^s_i$ is the pseudo gauge field. 
The explicit forms of $A_i^s$ and $v_{ij}$ for general displacement fields $\mathbf{u}$ can be found in \cite{wagner2022}, and the specific expressions for the strain profile \eqref{eq:straindef} are given below.
The spectrum of this system is  obtained from $\mathcal{H}_{TB}\psi = E\psi$, where the
scalar product $\left< \psi_1|\psi_2\right> = \int d^2x
\psi^{\dagger}_1 \psi_2$ is inherited from the original microscopic model.

The appearance of the spatial dependence in $v_{ij}$, and the associated term $\Gamma_i$, invalidate the common assumption that the effects of the strain can all be captured by an additional contribution to the vector potential.
On the other hand, such terms appear naturally for a Dirac fermion in a curved space of metric $g_{ij}$, coupled to a background $U(1)$ vector potential $\mathcal{A}_i$. This situation is described by the
Hamiltonian density \cite{wagner2022}
\begin{equation}
\mathcal{H} = i v_F e^i_a \gamma^a (\partial_i -\Omega_i-i\mathcal{A}_i) , \quad \Omega_i=-\frac{1}{2}e^a_i \d_j e^j_a -\frac{1}{2} \d_i \sqrt{g},
\label{Hdens}
\end{equation}
which defines the total Hamiltonian
\begin{equation}
H = \int d^2 x \sqrt{g} \psi^{\dagger}\mathcal{H} \psi ,\label{Hcurved}
\end{equation}
where $g=\mathrm{det}\,(g_{ij})$.
The important point is to realize is that 
$\mathcal{H}$ is Hermitian with respect to the scalar product $\int d^2x
\sqrt{g}\psi_1^{\dagger}\psi_2$, so that $\mathcal{H}$ cannot be directly compared to
$\mathcal{H}_{TB}$. The simplest way to allow the comparison is to redefine the curved space field
as $\psi = \tilde{\psi}g^{-1/4}$, so that we have a regular eigenvalue problem
\begin{equation}
\tilde{\mathcal{H}} \tilde{\psi} = i v_F e^i_a \gamma^a (\partial_i -\Omega_i-i\mathcal{A}_i-\frac{1}{2} \partial_i \sqrt{g})
\tilde{\psi} = E\tilde{\psi}
\end{equation}
with the standard scalar product. 
This matches with the Hamiltonian \eqref{HamTB} when we set
\begin{equation}
	e^i_a=v_{ia}, \quad \mathcal{A}_i=A_i+A^s_j \hat{v}_{ji}
 ,
\end{equation}  
where we define the inverse velocity matrix $\hat{v}_{ij}v_{jk}=\delta_{ik}$.
The corresponding induced magnetic field and  Gaussian curvature are
\begin{equation}
	\mathcal{B}=\frac{\epsilon^{ij} \d_i \mathcal{A}_j}{\sqrt{g}}, \qquad \mathcal{K}=\frac{\epsilon^{ij} \d_i \mathcal{\omega}_j}{\sqrt{g}}.
\end{equation} 

\section{Application to optimal strain}
\label{sec:unifcont}
In this Appendix, 
we present the derivation of the analytical results for a uniform pseudomagnetic field, which forms the basis of the discussion in Sec.~\ref{sec:uniform}.
For this, we utilize the general results from the continuum theory of the preceding App.~\ref{sec:cont},

In the continuum theory, the coupling configuration \eqref{eq:straindef} used in the main text corresponds to the standard triaxial strain field  \cite{guinea,wagner2022} 
\begin{equation}
\label{eq:u}
	\mathbf{u}=u_B\left(\begin{matrix}
		2 xy \\
		x^2-y^2 
	\end{matrix}\right),
\end{equation}  
coupled with a strength $\tilde{\beta}=\beta/(4u_B)$. 
Using the general expressions of Ref.~\cite{wagner2022},  the vielbein $e_a^i$ and pseudovector potential $A^s_i$ are then given by
\begin{equation}
    e_a^i=\left(\begin{matrix} 1-\frac{\beta}{4}y & - \frac{\beta}{4}x \\ - \frac{\beta}{4}x & 1+\frac{\beta}{4}y \end{matrix}\right), \quad \mathbf{A}^s=\frac{\beta}{2}\left(\begin{matrix} -  y \\  x\end{matrix}\right).
\end{equation}
Notably, this corresponds to constant values of the induced pseudomagnetic field and curvature,
\begin{equation}
\mathcal{B}=\beta, \quad 
\mathcal{K}=- \frac{\beta^2}{4}.
\end{equation}

In general, a Dirac fermion in a constant magnetic field $\mathcal{B}$ and a constant Gaussian curvature $\mathcal{K}$ has the spectrum
\cite{GGV92,pnuelli1994,wagner2022}
\begin{equation}
E_n = v_F \sign(n)\sqrt{2|n \mathcal{B}| + n^2 \mathcal{K}} 
.
\label{eq:Landau}
\end{equation}
For $\mathcal{K}>0$ this is the spectrum of the sphere \cite{GGV92}. For $\mathcal{K}<0$, which we encounter here, this expression gives
the discrete part of the spectrum only, which is constrained by $|n|< |\mathcal{B}|/|\mathcal{K}|$ \cite{GG08}. For larger
energies, there is a continuum of states. Furthermore, the degeneracy of these Landau levels is given by \cite{pnuelli1994} 
\begin{equation}
N_{e,n} = \frac{1}{2\pi} \int d^2x (|\mathcal{B}|+\frac{S_n}{2}\mathcal{K}) \sim 
\left(1+S_n\frac{\mathcal{K}}{2|\mathcal{B}|}\right)
,
\label{degeneracies}
\end{equation}
where $S_n=2|n|$ is the relativistic Wen-Zee shift. 
As we explain in the main text, these expressions coincide precisely with the exact results of the microscopic model in the optimally strained triangle configuration.

\section{Additional external magnetic field}
\label{sec:appunifmag}
In this Appendix, we provide the derivations for the analytical results in Sec.~\ref{sec:uniformmag}, where we consider the signatures of the hyperbolic geometry for an additionally applied external magnetic field. 

In the continuum theory, this magnetic field is obtained from a vector potential with $\epsilon^{ij}\d_i A_j= B$.
Mapped to the Dirac equation in curved space, the combined effective field takes the form
\begin{equation}
\label{eq:cm}
\mathcal{B}=\beta+\tilde{B},	
\end{equation}
where
\begin{align}
    \tilde{B}&=\frac{B}{\sqrt{g}} =
    B|1-\beta^2(x^2+y^2)/16|
\end{align}
acquires an induced spatial dependence.  
At the center of the sample, this effective magnetic field takes the combined value $\mathcal{B}=\beta + B$,  but at larger distances it reduces to the contribution from the pseudomagnetic field only, approaching $\mathcal{B}=\beta$ at a distance $r=4/\beta$, which happens to coincide with the finite system size obtained from the exact solution of the microscopic model. 

When one considers the effective theory for the fermion near the $K'$ point, the pseudomagnetic field induced by the strain switches sign \cite{guinea,zubkov2015,wagner2022},
\begin{equation}
\label{eq:cmp}
	\mathcal{B}=-\beta+\tilde{B},
\end{equation}
while the curvature remains the same. 
Consequently, when one applies both strain and an external magnetic field,
 the local Landau levels near the center of the sample take the valley-dependent energies
\begin{equation}
E^{\pm}_n = v_F \sign(n)\sqrt{2|n (\pm \beta+B)| - n^2 \frac{\beta^2}{4}} \label{eq:LandauV},	
\end{equation}
while far from the center they take the valley-independent form
\begin{equation}
E^{\pm}_n = v_F \sign(n)\sqrt{2|n \beta | - n^2 \frac{\beta^2}{4}} \label{eq:LandauV2}.	
\end{equation}
Taken altogether, these levels should therefore cover the interval between the two values in Eq.~\eqref{eq:LandauV}, which indeed agrees well with the numerical observations as discussed in the main text.

\section{Vortex state solutions}
\label{sec:vortcont}

In this Appendix, we derive the energy spectrum and wavefunctions of the vortex states considered in Sec.~\ref{sec:flux}.

\subsection{Vortex states near the $K$ point}

We initially ignore curvature effects, and derive solutions for vortex states in the pseudomagnetic field only $(B=0)$, where we assume $\beta>0$.
We define the complex coordinates
\begin{equation}
	z=\frac{1}{\sqrt{2}\ell_B}(x+iy), \quad \bar{z}=\frac{1}{\sqrt{2}\ell_B}(x-iy)
\end{equation}
with $\ell_B=1/\sqrt{\mathcal{B}}$, and $\mathcal{B}$ given by \eqref{eq:cm}. The effective Hamiltonian for a fermion near the $K$ point can then be written as
\begin{equation}
\label{eq:Hv}
	H_v=-\frac{v_F\sqrt{2}}{\ell_B} \left(\begin{matrix}
		0 & \hat{a}^\dagger \\
		\hat{a} & 0
	\end{matrix} \right),
\end{equation}
where we define the creation and annihilation operators
\begin{equation}
	\hat{a}=i {D}_{\bar z}, \quad \hat{a}^\dagger=i{D}_{ z}, 
\end{equation}
and the covariant derivatives
\begin{equation}
	D_z=\frac{\ell_B}{\sqrt{2}}(D_1-iD_2), \quad D_{\bar{z}}=\frac{\ell_B}{\sqrt 2}(D_1+iD_2),
\end{equation} 
with $D_i=\d_i -i \mathcal{A}_i$. 
If we insert a vortex with arbitrary flux $\varphi=\delta \phi_0$ ($|\delta| \leq 1/2$) at the center of the system, where $\phi_0=2\pi$ is the quantum flux, then the modified gauge field has the form 
\begin{equation}
\label{eq:avec}
	\mathcal{A}_x=-\frac{\mathcal{B}}{2}y- \delta\frac{y}{ r^2}, \quad \mathcal{A}_y=\frac{\mathcal{B}}{2}x + \delta\frac{x}{ r^2},
\end{equation}
with $r=\sqrt{x^2+y^2}$  the distance from the center. 
 We have the commutation relation
\begin{equation}
	[{D}_{z}, D_{\bar z}]=[\hat{a},\hat{a}^\dagger]=1. 
\end{equation}
Using the identity
\begin{multline}
	\hat{a}^\dagger \hat{a} (z^{\alpha_1} \bar{z}^{\alpha_2} e^{-z \bar z/2})=\\ \frac{(-\delta+2\alpha_1- 2 z\bar z)(-\delta-2\alpha_2)}{4 z \bar z}z^{\alpha_1} \bar{z}^{\alpha_2} e^{-z \bar z/2}
\end{multline}
we find that the Hamiltonian \eqref{eq:Hv} admits vortex states with energy
\begin{equation}
\label{eq:EvK}
	E^\lambda_{n+\delta}=\lambda v_F\sqrt{(2n+2\delta)\mathcal{B}}
.
\end{equation} 
The corresponding states are of the form
\begin{equation}
	\label{eq:wfarbi}
	\psi^m_{n+\delta}=\left(\begin{matrix}
		|n+\delta\rangle_m \\
		-\lambda |n-1+ \delta\rangle_m
	\end{matrix}\right) \qquad (n\geq 1)
 ,
\end{equation}
where the states in the components satisfy 
\begin{equation}
	\label{eq:aa1}
	\hat{a}^\dagger \hat{a}	|n+\delta\rangle_m=(n+\delta)|n+\delta\rangle_m \qquad (n\geq 0)
 .
\end{equation}
The explicit coordinate form of the normalizable wavefunction is given by
\begin{equation}
\label{eq:wfvK}
	|n+\delta\rangle_m=\sum_{j=0}^{m }  c_j z^{\frac{\delta}{2}+j}\bar{z}^{\frac{\delta}{2}+n-m+j}e^{-z \bar{z}/2} \quad (0 \leq m \leq n
 )
 ,
\end{equation}
where the coefficients $c_i$ satisfy the recursion relation
\begin{equation}
	j  \left[-\delta-(n-m+j)\right]c_j=(m+1-j)c_{j-1}
 .
\end{equation}
There are $n+1$ normalizable states that satisfy \eqref{eq:aa1}, which translate into $n$ eigenstates
 with wavefunction \eqref{eq:wfarbi}. Each of these has the same energy, given by Eq.~\eqref{eq:EvK}.

In addition to these states, we still have the usual Landau levels at energy $E^\lambda_n=\lambda v_F\sqrt{2\mathcal{B}n} $. These are described by the family of wavefunctions
\begin{equation}
	\psi^f_n=\left(\begin{matrix}
		|n\rangle_f \\
		-\lambda |n-1\rangle_f
	\end{matrix}\right), \quad \lambda=\pm 1
 ,
\end{equation}
where the wavefunction 
\begin{equation}
	|0\rangle_f=f(z)(z \bar{z})^{-\delta/2}e^{-z \bar z/2}
\end{equation}
of the zeroth Landau level
satisfies $a|0\rangle_f=0$ and $f(z)$ is a holomorphic function of $z$,
while the higher Landau levels are obtained from the definition
\begin{equation}
	|n\rangle_f=\frac{1}{\sqrt{n !}} \left(\hat{a}^\dagger\right)^n|0\rangle_f
 .
\end{equation}
\subsection{Vortex state near the $K'$ point}

Near the $K'$ point, the pseudomagnetic field switches sign in the Dirac Hamiltonian, but the external magnetic field is preserved \cite{zubkov2015}. While the gauge potential
\begin{equation}
	\mathcal{A}_x=-\frac{\mathcal{B}}{2}y- \delta\frac{y}{ r^2}, \quad \mathcal{A}_y=\frac{\mathcal{B}}{2}x + \delta\frac{x}{ r^2}
\end{equation}
reads the same as Eq.~\eqref{eq:avec}, $\mathcal{B}$  is then given by \eqref{eq:cmp}. As $\beta>0$ now implies  $\mathcal{B}<0$,
the roles of creation and annihilation operators become interchanged. 
Using the identity
\begin{multline}
	\hat{a} \hat{a}^\dagger (z^{\alpha_1} \bar{z}^{\alpha_2} e^{-z \bar z/2})=\\ \frac{(-\delta-2\alpha_2+ 2 z\bar z)(-\delta+2\alpha_1)}{4 z \bar z}z^{\alpha_1} \bar{z}^{\alpha_2} e^{z \bar z/2}
 ,
\end{multline}
we can repeat the calculation to obtain the eigenenergies 
\begin{equation}
\label{eq:EvKp}
	E'^\lambda_{n-\delta}=\lambda v_F\sqrt{(2n-2\delta)|\mathcal{B}|}
\end{equation}
and states 
\begin{equation}
	\label{eq:wfarbi2}
	\psi'^{m}_{n-\delta}=\frac{1}{\sqrt{2}}\left(\begin{matrix}
		|n-1-\delta\rangle_m \\
		-\lambda |n- \delta\rangle_m
	\end{matrix}\right) \qquad (n\geq 1),
\end{equation}
where the wavefunction  satisfies 
\begin{equation}
	\label{eq:aa2}
	\hat{a} \hat{a}^\dagger	|n-\delta\rangle_m=(n-\delta)|n-\delta\rangle_m \quad (n\geq 0).
\end{equation}
The explicit form of the normalizable wavefunctions is now given by
\begin{equation}
\label{eq:wfvKp}
	|n+\delta\rangle_m=\sum_{j=0}^{m } c_j \bar{z}^{-\frac{\delta}{2}+j}z^{-\frac{\delta}{2}+n-m+j}e^{-z \bar{z}/2}   \quad (0 \leq m \leq n
 ),
\end{equation}
where the coefficients $c_i$ now satisfy the recursion relation
\begin{equation}
	-j  (-\delta+(n-m+j))c_j=(m+1-j)c_{j-1}
 ,
\end{equation}
There are again $n+1$ states that satisfy \eqref{eq:aa1}, which translate into $n$ eigenstates
with wavefunction \eqref{eq:wfarbi2}, each having the energy given in Eq.~\eqref{eq:EvKp}. 

Again, we still have the usual Landau levels at energies $E'^\lambda_n=\lambda v_F\sqrt{2 n |\mathcal{B}|} $, whose wavefunctions
\begin{equation}
	\psi'^f_n=\frac{1}{\sqrt{2}}\left(\begin{matrix}
		|n-1\rangle_f \\
		-\lambda |n\rangle_f
	\end{matrix}\right), \quad \lambda=\pm 1
\end{equation}
now follow from the definitions
\begin{equation}
	|n\rangle_f=\frac{1}{\sqrt{n !}} \left(\hat{a}\right)^n|0\rangle_f
\end{equation}
and 
\begin{equation}
	|0\rangle_f=f(\bar{z})(z \bar{z})^{\delta/2}e^{-z \bar z/2}, 
\end{equation}
while $f(\bar z)$ now is an anti-holomorphic function of $\bar z$.

In the case $\delta=\pm\frac{1}{2}$ of a half-integer vortex, the spectra of the vortex states near the $K$ and $K'$ points are identical. However, for general values of $\delta$ this degeneracy is lifted. Therefore, between  Landau levels $n$ and $n+1$  there are $n+1$ vortex states with energy $E_{n+1-\delta}$ that come from the $K'$ valley,  and  $n$ such states with energy $E'_{n+\delta}$ that come from the $K$ valley.  From the explicit forms of their wavefunction 
\eqref{eq:wfvK} and \eqref{eq:wfvKp}, we also see that with increasing $n$ they contain higher powers of $z$ and $\bar{z}$, so that the expectation value  $\langle r \rangle=\ell\langle\sqrt{z\bar z}\rangle$ of their radius increases. At small $n$, the vortex states are tightly bound to the vortex. 

\subsection{Curvature effects}
Finally, we consider the additional effects of a constant curvature $\mathcal{K}$ and an additional uniform magnetic field in $\mathcal{B}$. Following \cite{pnuelli1994}, one can show that 
\begin{align}
&\left(H_v(\mathcal{B},\mathcal{K})\right)^2
\nonumber
\\
&=\left(\begin{matrix}H_s(\mathcal{A}^{-}, \mathcal{K})-\mathcal{B}+\frac{\mathcal{K}}{2} & 0 \\ 0 & H_s(\mathcal{A}^{+}, \mathcal{K})+\mathcal{B}+\frac{\mathcal{K}}{2} \end{matrix} \right)
\label{eq:H2}
\end{align} 
where $H_s(\mathcal{A}, \mathcal{K})$ is the Schr{\"o}dinger Hamiltonian of a non-relativistic particle on a Riemann surface with a constant curvature, which couples to a background vector potential $\mathcal{A}$ as defined in Ref.~\cite{pnuelli1994}. The gauge potential $\mathcal{A}^+$ gives a vortex flux $\delta \phi_0$ at the origin and a constant magnetic field $\mathcal{B}+\frac{\mathcal{K}}{2}$. Similarly $\mathcal{A}^-$ gives a vortex flux $\delta \phi_0$ at the origin and a constant magnetic field $\mathcal{B}-\frac{\mathcal{K}}{2}$. The eigenvalue of \eqref{eq:H2} that is consistent with the energy vortex states \eqref{eq:EvK} at $\mathcal{K}=0$ is given by 
\begin{equation}
\label{eq:EvKcur2}
	(E^\lambda_{n+\delta}(\mathcal{K}))^2=2v_F^2(|n|+\delta)|\mathcal{B}|+v_F^2(|n|+\delta)^2 \mathcal{K},
\end{equation}
which implies 
\begin{equation}
\label{eq:EvKcur}
	E^\lambda_{n+\delta}(\mathcal{K})=\lambda v_F\sqrt{2(|n|+\delta)|\mathcal{B}|+(|n|+\delta)^2 \mathcal{K}} , \quad \lambda=\pm 1.
\end{equation}
These expressions are
consistent with Eqs.~\eqref{eq:EvK} and \eqref{eq:EvKp} in the limit  $\mathcal{K}=0$, and with Eq.~\eqref{eq:Landau} in the limit $\delta=0$.\\   

Similarly, the energy of vortex states near the $K'$ point including the constant curvature is
\begin{equation}
\label{eq:EvKpcur}
	E'^\lambda_{n-\delta}(\mathcal{K})=\lambda v_F\sqrt{2(|n|-\delta)|\mathcal{B}|+(|n|-\delta)^2 \mathcal{K}} , \quad \lambda=\pm 1.
\end{equation}

\begin{figure}
    \centering    \includegraphics[width=\columnwidth]{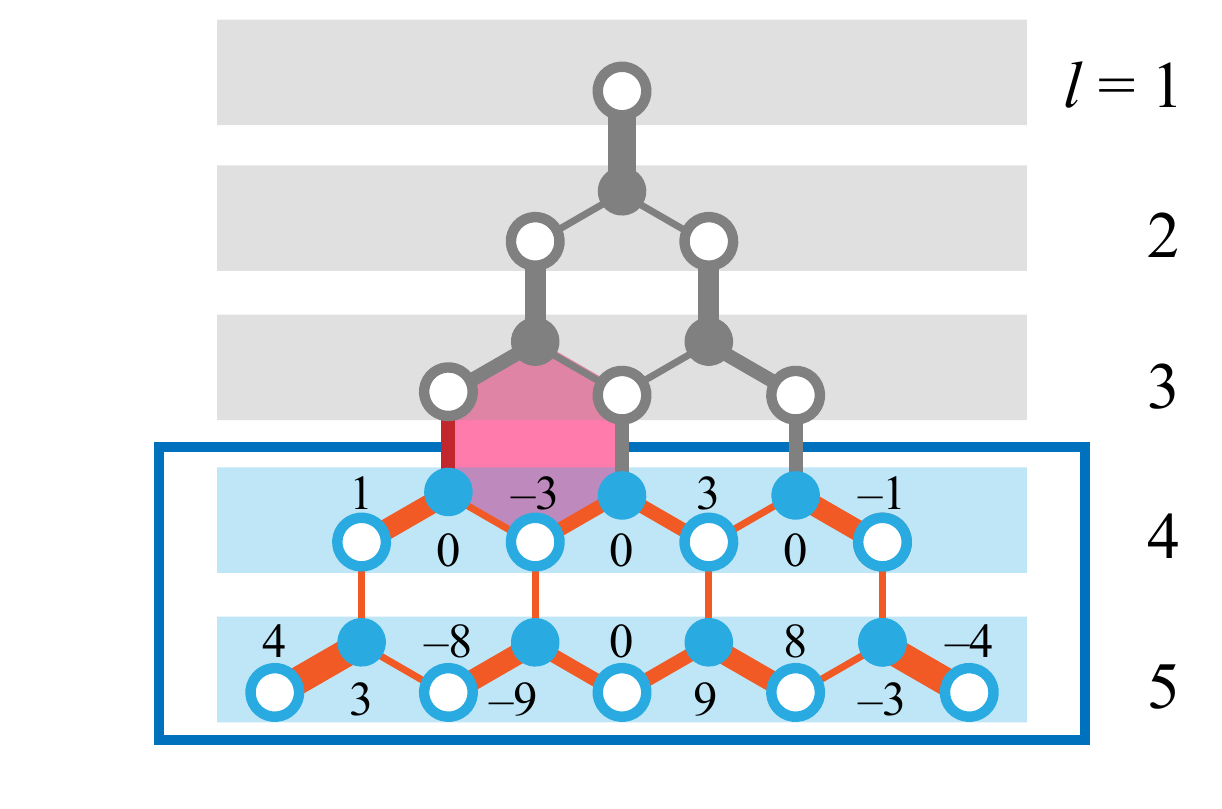}
    \caption{Division of an optimally strained triangle into zigzag chains labelled by $l$, as used for the construction of exact pLL states in App.~\ref{sec:pllapp}. The amplitudes specify a state in the first pLL for system size $N=5$. This state is supported by the blue trapezoidal region, and hence is unaffected by the indicated flux vortex, which is generated by a phase shift of the deep red coupling.}
    \label{fig9}
\end{figure}

\section{Explicit construction of pseudo-Landau level states in the optimal strain configuration}
\label{sec:pllapp}
Here, we first recapitulate the construction of exact pLL states in the microscopic model with the optimal strain configuration \eqref{eq:straindef},  and then confirm the adaptation to systems with a flux vortex 
discussed in Sec.~\ref{sec:fluxa}. 
For this, we bring the results from Ref.~\cite{rachel2016} into an explicit form, where we are guided by Refs.~\cite{cas2014,exp5}. This is achieved by dividing the triangular system into $N$ zigzag chains of alternating A and B sites, running parallel to one of the edges, as illustrated in Fig.~\ref{fig9}. Along the $l$th chain, we denote the amplitudes on the A sites as 
    $A_{l,m}$ ($m=1\ldots l$)
and on the B sites as  $B_{l,m}$ ($m=1\ldots l-1$). As indicated, each chain contains one more A site than B sites.
Furthermore, along this chain the couplings alternate as
\begin{equation}
    \frac{3t}{N} (l-1,1,l-2,2,\ldots,2, l-2,1, l-1),
\end{equation}
while the coupling between chains $l$ and $l+1$ is given by $3t (N-l)/N$.
This suggests to set $t=N/3$ so that all couplings are integers, which we adopt from here on. 
The eigenvalue equation can then explicitly be written as
\begin{align}
EA_{l,m}&=(l-m)B_{l,m}+(m-1)B_{l,m-1}+(N-l)B_{l+1,m},
\label{eq:tbexpl1}
\\
EB_{l,m}&=(l-m)A_{l,m}+mA_{l,m+1}+(N-l+1)A_{l-1,m}.
\label{eq:tbexpl2}
\end{align}

We next construct the exact zero-energy states of the system. Because of the chiral symmetry and the imbalance of sites on both sublattices, these are all localized on the A sites. 
Throughout the system, we then have to fulfill the condition 
\begin{equation}
    0=(l-m)A_{l,m}+mA_{l,m+1}+(N-l+1)A_{l-1,m},
\end{equation}
which provides a solution of Eqs.~\eqref{eq:tbexpl1}, \eqref{eq:tbexpl2}
with vanishing amplitude $B_{l,m}$.
We find the explicit solutions
\begin{equation}
\label{eq:exactstates}
    A_{l,m}^{N,0,k}=(-1)^{l+m}{N-k\choose m-1}{k-1\choose N-l},
\end{equation}
where $r\choose s$ denotes binomial coefficients, $N$ denotes the system size, and $k=1,\ldots, N$ labels the different states.

Importantly, each of these states has finite amplitudes on only $k$ chains adjacent to the edge, covering a trapezoidal region with $N-k+1\leq l\leq N$. Such a state remains an exact solution even when we remove or modify parts of the system away from its support.

Starting from these states, the finite-energy pLLs can be constructed recursively by setting \cite{rachel2016}
\begin{align}
A_{l,m}^{N,n+1,k}&=(l-m)
A_{l-1,m}^{N-1,n,k}
+(m-1)A_{l-1,m-1}^{N-1,n,k}
\nonumber\\
&+(N-l)A_{l,m}^{N-1,n,k},    \\
B_{l,m}^{N,n+1,k}&=E_{N,n+1}A_{l-1,m}^{N-1,n,k}.
\end{align}
This automatically fulfills Eq.~\eqref{eq:tbexpl1}, while consistency with Eq.~\eqref{eq:tbexpl2} enforces the quantization condition $E_{N,n}=\mathrm{sgn}\,(n)\,\sqrt{|n|(2N-|n|)}$.
Furthermore, each level contains one fewer state, giving the  degeneracy $D_n=N-|n|$, with $-N<n<N$. 
As $\sum_n D_n=N^2$ equals the number of sites on the triangle, this exhausts all states in the Hilbert space of the system.

Starting the recursion with states from Eq.~\eqref{eq:exactstates}, we see that each index $k$ delivers a sequence of pLL states with $|n|< k$. Just as the initializing zero-energy state, these states furthermore have a finite support in a trapezoidal region of width $k$, as we illustrate by an example in Fig.~\ref{fig9}. 
These states, therefore, remain exact solutions when the system is truncated to these regions,
and can be directly transferred to a triangular system with an additional flux vortex placed outside the trapezoid, as again illustrated in the Figure. 
This is our main result from this construction.

For a trapezoid of width $N_i$, as indicated in Fig.~\ref{fig5},
we then obtain a sequence of pLLs with degeneracy $N_i-|n|$, $|n|<N_i$, and these also provide solutions for the system with the flux vortex.  In the subsystem crossed by the string connecting the vortex to the boundary, this is subject to a gauge transformation $\psi_l\to \exp^{-i\varphi}\psi_l$ for all sites below the string. 
The remaining states become midgap states, as described in the main text.

\begin{figure}[t]
\includegraphics[width=\linewidth]{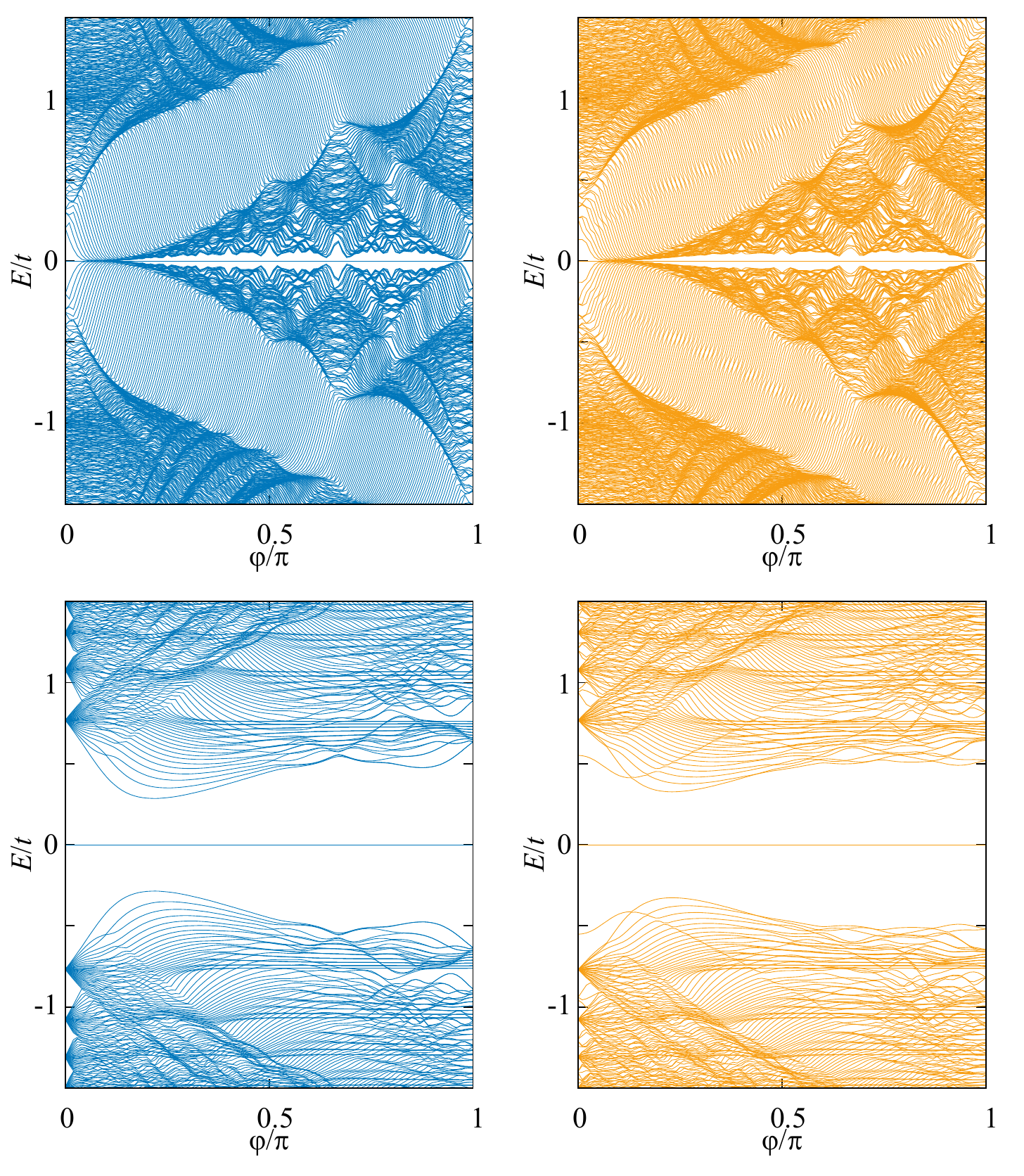}
\caption{\label{fig10}
Magnetic-field dependence of the energies in a triangular system of size $N=30$, without any vortices (left), and with a half-integer vortex placed into the center (right). In the top panels, the system is pristine, in the bottom panels it is optimally strained.}
 \end{figure}

 \begin{figure}[t]
\includegraphics[width=\columnwidth]{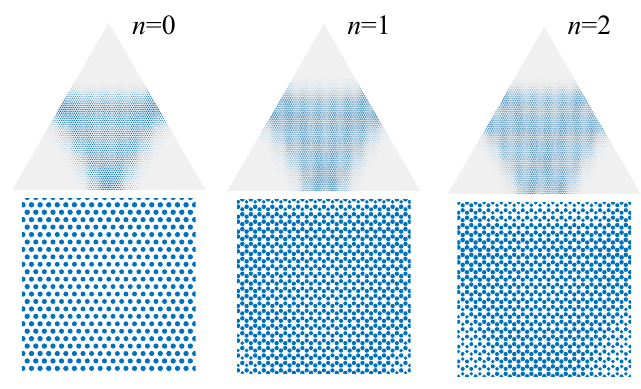}
\caption{\label{fig11}
Support of the lowest pLLs in the optimally strained triangle without a flux vortex ($N=90$). The bottom panels zoom into the central region.}
\end{figure}

 \begin{figure}[t]
\includegraphics[width=\columnwidth]{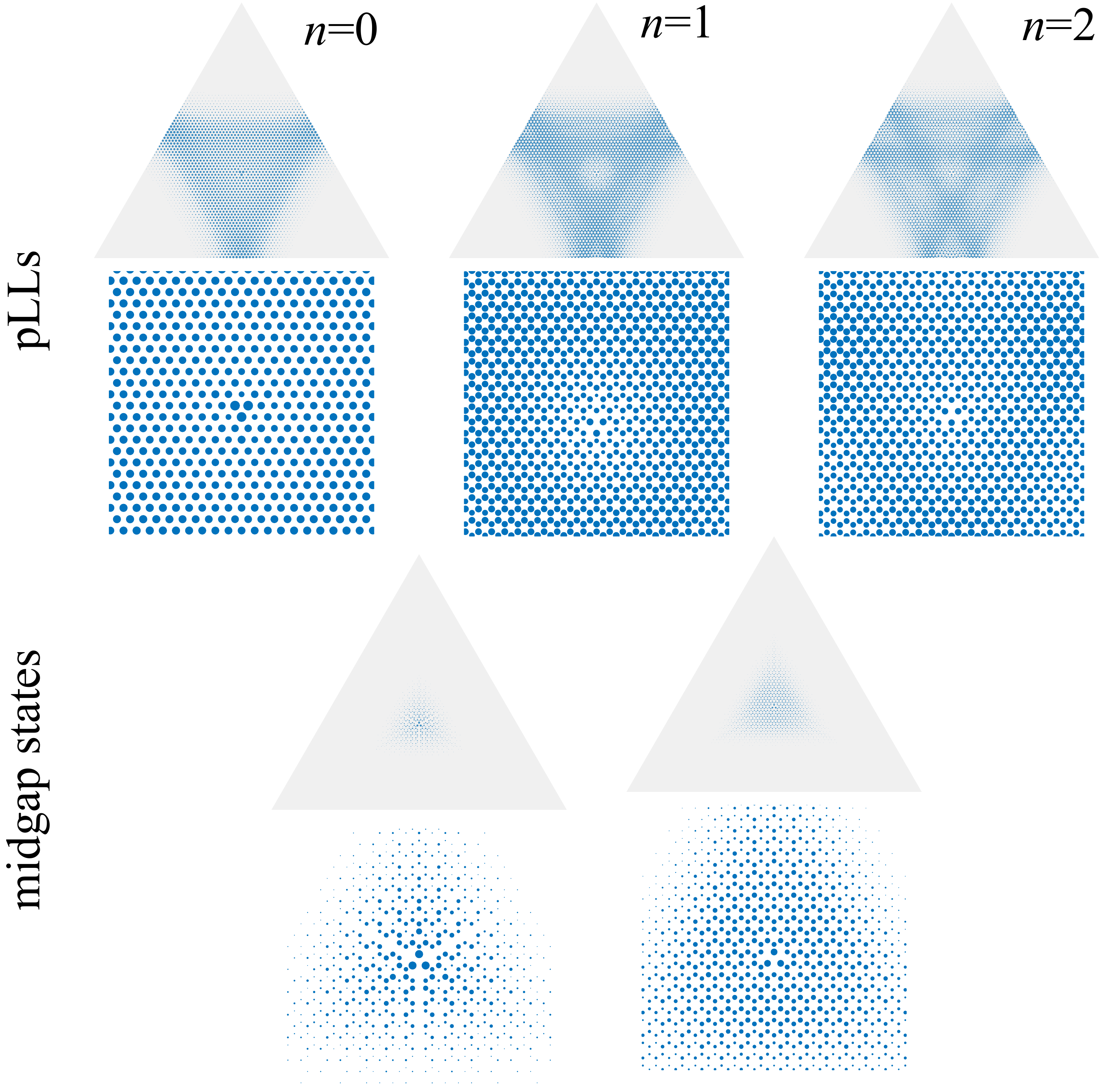}
\caption{\label{fig12}
Support of the lowest pLLs and midgap states in the optimally strained configuration with a central half-integer flux vortex ($N=90$).} 
\end{figure}

\begin{figure}[t]
  \centering
    \includegraphics[width=\columnwidth]{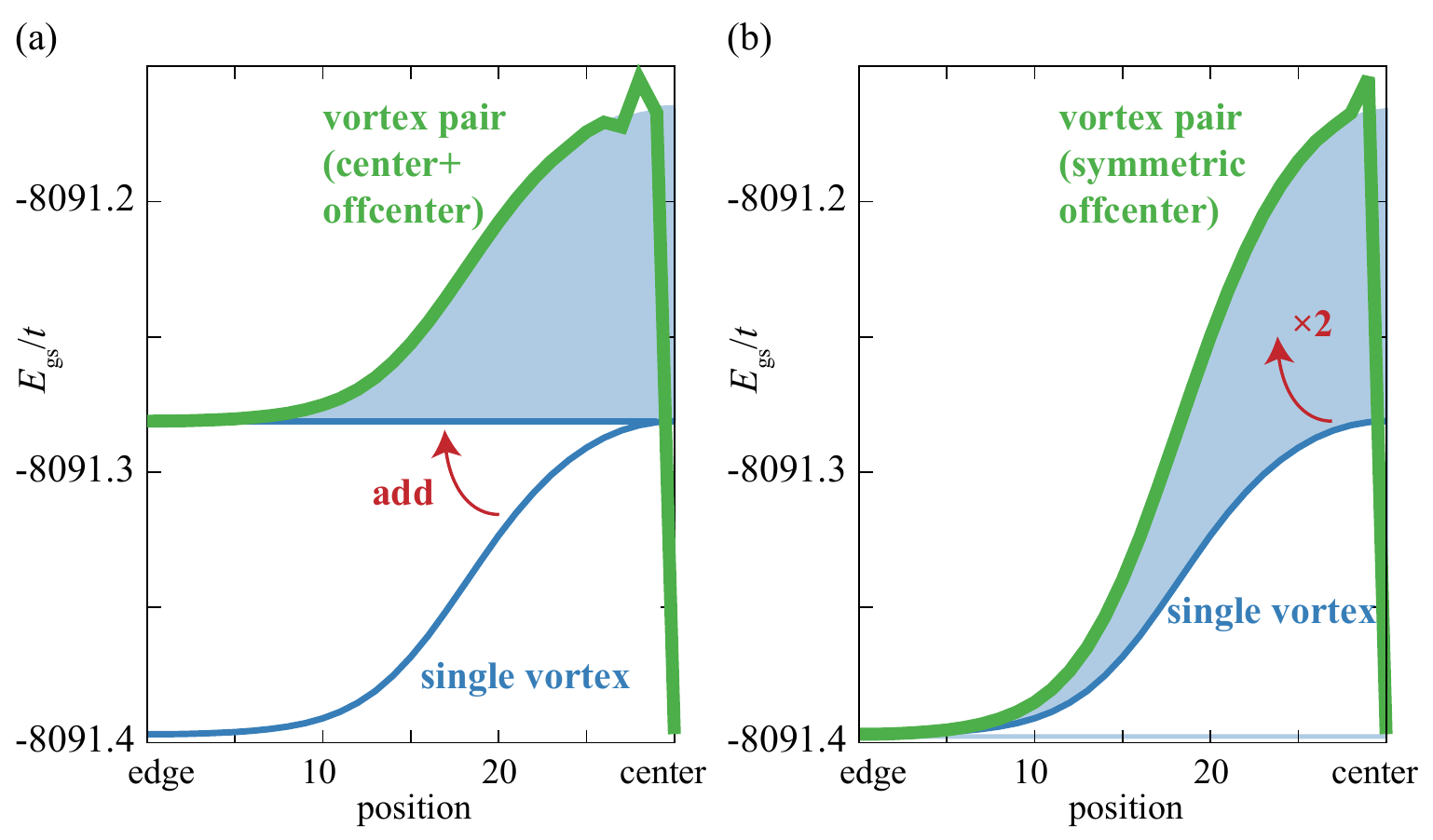}
  \caption{(a) Determination of the effective vortex interaction energy in Fig.~\ref{fig8}(c) from the ground states of  sectors in different vortex configurations, amounting to the difference between the shaded and green curves
  in accordance with Eq.~\eqref{eq:vint}.
  This interaction energy is consistent with that obtained by displacing both vortices symmetrically from the center, giving the ground state energies of panel (b). This data is for optimally strained triangles of size $N=90$.
}  
  \label{fig13}
\end{figure}

\begin{figure}[b]
  \centering
  \includegraphics[width=\columnwidth]{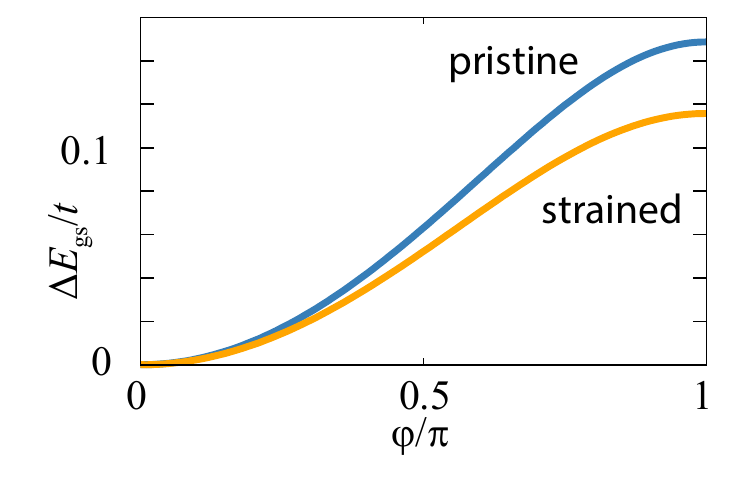}
  \caption{Excess energy in the many-body ground state of as in triangular Kitaev honeycomb system, but generalized to a central vortex of arbitrary flux. The system size is $N=90$, and the result in the pristine (unstrained) configuration is compared to that in the optimally strained configuration. }\label{fig14}
\end{figure}

\section{Supplementary numerical results}
\label{sec:addnum}

Here we provide additional numerical results that illustrate further points in the main text.

Throughout, we focussed on the system with pseudo-Landau levels generated by strain. 
As shown in the top panels of Fig.~\ref{fig10}, a noticeably different picture emerges when we completely replace the pseudomagnetic field by a real magnetic field. For the resulting magnetic LLs, the degeneracy does not saturate but increases as the magnetic field increases, until the familiar Hofstadter butterfly picture emerges. The gaps are then always filled by (edge) states that are in the process of crossing over from one LL to another LL. A half-flux vortex gives rise to a weak modulation of this pattern within the gaps, but not to the extent that individual midgap states could be identified.
The bottom panels depict the magnetic-field dependence in the optimally strained system. The pLLs then quickly broaden, as described in detail in Sec.~\ref{sec:uniformmag}. Noticeably, a large gap remains around the zeroth pLL.

In Sec.~\ref{sec:uniform} we remarked that the lowest pLLs in the optimally strained system fill out the central triangle in which the system is locally not gapped. This feature is illustrated in Fig.~\ref{fig11}. As shown in Fig.~\ref{fig12}, the pLLs in the system with a central half-integer flux vortex are depleted around this vortex. This weight is taken up the midgap states. 

In Sec.~\ref{sec:interactions} we considered ground state energies in different vortex configurations of half-integer fluxes, as relevant for the application to the Kitaev honeycomb system. As shown in Fig.~\ref{fig13}, spatially well-separated vortices provide independent excess energies. The figure then further illustrates how these can be combined to define the effective interaction energy \eqref{eq:vint}. 

Finally, in Fig.~\ref{fig14} we show the ground state if the system supported a central vortex of general flux $\varphi$, making use that this preserves the chiral symmetry so that many-body energies can still be formally obtained from the single-particle energy pairs. Alternatively, this result could be interpreted as a ground-state energy of noninteracting electrons in an analogous graphene-like system, evaluated  per spin  at half-filling.

\bibliography{strainHC}

\begin{thebibliography}{49}%
\makeatletter
\providecommand \@ifxundefined [1]{%
 \@ifx{#1\undefined}
}%
\providecommand \@ifnum [1]{%
 \ifnum #1\expandafter \@firstoftwo
 \else \expandafter \@secondoftwo
 \fi
}%
\providecommand \@ifx [1]{%
 \ifx #1\expandafter \@firstoftwo
 \else \expandafter \@secondoftwo
 \fi
}%
\providecommand \natexlab [1]{#1}%
\providecommand \enquote  [1]{``#1''}%
\providecommand \bibnamefont  [1]{#1}%
\providecommand \bibfnamefont [1]{#1}%
\providecommand \citenamefont [1]{#1}%
\providecommand \href@noop [0]{\@secondoftwo}%
\providecommand \href [0]{\begingroup \@sanitize@url \@href}%
\providecommand \@href[1]{\@@startlink{#1}\@@href}%
\providecommand \@@href[1]{\endgroup#1\@@endlink}%
\providecommand \@sanitize@url [0]{\catcode `\\12\catcode `\$12\catcode
  `\&12\catcode `\#12\catcode `\^12\catcode `\_12\catcode `\%12\relax}%
\providecommand \@@startlink[1]{}%
\providecommand \@@endlink[0]{}%
\providecommand \url  [0]{\begingroup\@sanitize@url \@url }%
\providecommand \@url [1]{\endgroup\@href {#1}{\urlprefix }}%
\providecommand \urlprefix  [0]{URL }%
\providecommand \Eprint [0]{\href }%
\providecommand \doibase [0]{https://doi.org/}%
\providecommand \selectlanguage [0]{\@gobble}%
\providecommand \bibinfo  [0]{\@secondoftwo}%
\providecommand \bibfield  [0]{\@secondoftwo}%
\providecommand \translation [1]{[#1]}%
\providecommand \BibitemOpen [0]{}%
\providecommand \bibitemStop [0]{}%
\providecommand \bibitemNoStop [0]{.\EOS\space}%
\providecommand \EOS [0]{\spacefactor3000\relax}%
\providecommand \BibitemShut  [1]{\csname bibitem#1\endcsname}%
\let\auto@bib@innerbib\@empty
\bibitem [{\citenamefont {Mermin}(1979)}]{rev1}%
  \BibitemOpen
  \bibfield  {author} {\bibinfo {author} {\bibfnamefont {N.~D.}\ \bibnamefont
  {Mermin}},\ }\bibfield  {title} {\bibinfo {title} {The topological theory of
  defects in ordered media},\ }\href
  {https://doi.org/10.1103/RevModPhys.51.591} {\bibfield  {journal} {\bibinfo
  {journal} {Rev. Mod. Phys.}\ }\textbf {\bibinfo {volume} {51}},\ \bibinfo
  {pages} {591} (\bibinfo {year} {1979})}\BibitemShut {NoStop}%
\bibitem [{\citenamefont {Hasan}\ and\ \citenamefont {Kane}(2010)}]{rev2}%
  \BibitemOpen
  \bibfield  {author} {\bibinfo {author} {\bibfnamefont {M.~Z.}\ \bibnamefont
  {Hasan}}\ and\ \bibinfo {author} {\bibfnamefont {C.~L.}\ \bibnamefont
  {Kane}},\ }\bibfield  {title} {\bibinfo {title} {Colloquium: Topological
  insulators},\ }\href {https://doi.org/10.1103/RevModPhys.82.3045} {\bibfield
  {journal} {\bibinfo  {journal} {Rev. Mod. Phys.}\ }\textbf {\bibinfo {volume}
  {82}},\ \bibinfo {pages} {3045} (\bibinfo {year} {2010})}\BibitemShut
  {NoStop}%
\bibitem [{\citenamefont {Qi}\ and\ \citenamefont {Zhang}(2011)}]{rev3}%
  \BibitemOpen
  \bibfield  {author} {\bibinfo {author} {\bibfnamefont {X.-L.}\ \bibnamefont
  {Qi}}\ and\ \bibinfo {author} {\bibfnamefont {S.-C.}\ \bibnamefont {Zhang}},\
  }\bibfield  {title} {\bibinfo {title} {Topological insulators and
  superconductors},\ }\href {https://doi.org/10.1103/RevModPhys.83.1057}
  {\bibfield  {journal} {\bibinfo  {journal} {Rev. Mod. Phys.}\ }\textbf
  {\bibinfo {volume} {83}},\ \bibinfo {pages} {1057} (\bibinfo {year}
  {2011})}\BibitemShut {NoStop}%
\bibitem [{\citenamefont {Wilczek}(1982)}]{anyon0}%
  \BibitemOpen
  \bibfield  {author} {\bibinfo {author} {\bibfnamefont {F.}~\bibnamefont
  {Wilczek}},\ }\bibfield  {title} {\bibinfo {title} {Quantum mechanics of
  fractional-spin particles},\ }\href
  {https://doi.org/10.1103/PhysRevLett.49.957} {\bibfield  {journal} {\bibinfo
  {journal} {Phys. Rev. Lett.}\ }\textbf {\bibinfo {volume} {49}},\ \bibinfo
  {pages} {957} (\bibinfo {year} {1982})}\BibitemShut {NoStop}%
\bibitem [{\citenamefont {Stern}(2008)}]{anyon1}%
  \BibitemOpen
  \bibfield  {author} {\bibinfo {author} {\bibfnamefont {A.}~\bibnamefont
  {Stern}},\ }\bibfield  {title} {\bibinfo {title} {Anyons and the quantum
  {Hall} effect-a pedagogical review},\ }\href
  {https://doi.org/https://doi.org/10.1016/j.aop.2007.10.008} {\bibfield
  {journal} {\bibinfo  {journal} {Ann. Phys. (NY)}\ }\textbf {\bibinfo {volume}
  {323}},\ \bibinfo {pages} {204} (\bibinfo {year} {2008})}\BibitemShut
  {NoStop}%
\bibitem [{\citenamefont {Arovas}\ \emph {et~al.}(1985)\citenamefont {Arovas},
  \citenamefont {Schrieffer}, \citenamefont {Wilczek},\ and\ \citenamefont
  {Zee}}]{anyon2}%
  \BibitemOpen
  \bibfield  {author} {\bibinfo {author} {\bibfnamefont {D.~P.}\ \bibnamefont
  {Arovas}}, \bibinfo {author} {\bibfnamefont {R.}~\bibnamefont {Schrieffer}},
  \bibinfo {author} {\bibfnamefont {F.}~\bibnamefont {Wilczek}},\ and\ \bibinfo
  {author} {\bibfnamefont {A.}~\bibnamefont {Zee}},\ }\bibfield  {title}
  {\bibinfo {title} {Statistical mechanics of anyons},\ }\href
  {https://doi.org/https://doi.org/10.1016/0550-3213(85)90252-4} {\bibfield
  {journal} {\bibinfo  {journal} {Nucl. Phy. B}\ }\textbf {\bibinfo {volume}
  {251}},\ \bibinfo {pages} {117} (\bibinfo {year} {1985})}\BibitemShut
  {NoStop}%
\bibitem [{\citenamefont {Wilczek}(1990)}]{anyon3}%
  \BibitemOpen
  \bibfield  {author} {\bibinfo {author} {\bibfnamefont {F.}~\bibnamefont
  {Wilczek}},\ }\href {https://doi.org/10.1142/0961} {\emph {\bibinfo {title}
  {Fractional Statistics and Anyon Superconductivity}}}\ (\bibinfo  {publisher}
  {World Scientific},\ \bibinfo {year} {1990})\BibitemShut {NoStop}%
\bibitem [{\citenamefont {Wen}(1989)}]{WenFC1}%
  \BibitemOpen
  \bibfield  {author} {\bibinfo {author} {\bibfnamefont {X.~G.}\ \bibnamefont
  {Wen}},\ }\bibfield  {title} {\bibinfo {title} {Vacuum degeneracy of chiral
  spin states in compactified space},\ }\href
  {https://doi.org/10.1103/PhysRevB.40.7387} {\bibfield  {journal} {\bibinfo
  {journal} {Phys. Rev. B}\ }\textbf {\bibinfo {volume} {40}},\ \bibinfo
  {pages} {7387} (\bibinfo {year} {1989})}\BibitemShut {NoStop}%
\bibitem [{\citenamefont {Wen}\ and\ \citenamefont {Niu}(1990)}]{WenFC2}%
  \BibitemOpen
  \bibfield  {author} {\bibinfo {author} {\bibfnamefont {X.~G.}\ \bibnamefont
  {Wen}}\ and\ \bibinfo {author} {\bibfnamefont {Q.}~\bibnamefont {Niu}},\
  }\bibfield  {title} {\bibinfo {title} {Ground-state degeneracy of the
  fractional quantum {Hall} states in the presence of a random potential and on
  high-genus {Riemann} surfaces},\ }\href
  {https://doi.org/10.1103/PhysRevB.41.9377} {\bibfield  {journal} {\bibinfo
  {journal} {Phys. Rev. B}\ }\textbf {\bibinfo {volume} {41}},\ \bibinfo
  {pages} {9377} (\bibinfo {year} {1990})}\BibitemShut {NoStop}%
\bibitem [{\citenamefont {Chen}\ \emph {et~al.}(2014)\citenamefont {Chen},
  \citenamefont {Kee},\ and\ \citenamefont {Kim}}]{FCspinliquid}%
  \BibitemOpen
  \bibfield  {author} {\bibinfo {author} {\bibfnamefont {G.}~\bibnamefont
  {Chen}}, \bibinfo {author} {\bibfnamefont {H.-Y.}\ \bibnamefont {Kee}},\ and\
  \bibinfo {author} {\bibfnamefont {Y.~B.}\ \bibnamefont {Kim}},\ }\bibfield
  {title} {\bibinfo {title} {Fractionalized charge excitations in a spin liquid
  on partially filled pyrochlore lattices},\ }\href
  {https://doi.org/10.1103/PhysRevLett.113.197202} {\bibfield  {journal}
  {\bibinfo  {journal} {Phys. Rev. Lett.}\ }\textbf {\bibinfo {volume} {113}},\
  \bibinfo {pages} {197202} (\bibinfo {year} {2014})}\BibitemShut {NoStop}%
\bibitem [{\citenamefont {Ivanov}(2001)}]{nonabelianvortex}%
  \BibitemOpen
  \bibfield  {author} {\bibinfo {author} {\bibfnamefont {D.~A.}\ \bibnamefont
  {Ivanov}},\ }\bibfield  {title} {\bibinfo {title} {Non-abelian statistics of
  half-quantum vortices in $\mathit{p}$-wave superconductors},\ }\href
  {https://doi.org/10.1103/PhysRevLett.86.268} {\bibfield  {journal} {\bibinfo
  {journal} {Phys. Rev. Lett.}\ }\textbf {\bibinfo {volume} {86}},\ \bibinfo
  {pages} {268} (\bibinfo {year} {2001})}\BibitemShut {NoStop}%
\bibitem [{\citenamefont {Feiguin}\ \emph {et~al.}(2007)\citenamefont
  {Feiguin}, \citenamefont {Trebst}, \citenamefont {Ludwig}, \citenamefont
  {Troyer}, \citenamefont {Kitaev}, \citenamefont {Wang},\ and\ \citenamefont
  {Freedman}}]{Feiguin2007}%
  \BibitemOpen
  \bibfield  {author} {\bibinfo {author} {\bibfnamefont {A.}~\bibnamefont
  {Feiguin}}, \bibinfo {author} {\bibfnamefont {S.}~\bibnamefont {Trebst}},
  \bibinfo {author} {\bibfnamefont {A.~W.~W.}\ \bibnamefont {Ludwig}}, \bibinfo
  {author} {\bibfnamefont {M.}~\bibnamefont {Troyer}}, \bibinfo {author}
  {\bibfnamefont {A.}~\bibnamefont {Kitaev}}, \bibinfo {author} {\bibfnamefont
  {Z.}~\bibnamefont {Wang}},\ and\ \bibinfo {author} {\bibfnamefont {M.~H.}\
  \bibnamefont {Freedman}},\ }\bibfield  {title} {\bibinfo {title} {Interacting
  anyons in topological quantum liquids: The golden chain},\ }\href
  {https://doi.org/10.1103/PhysRevLett.98.160409} {\bibfield  {journal}
  {\bibinfo  {journal} {Phys. Rev. Lett.}\ }\textbf {\bibinfo {volume} {98}},\
  \bibinfo {pages} {160409} (\bibinfo {year} {2007})}\BibitemShut {NoStop}%
\bibitem [{\citenamefont {Kitaev}(2006)}]{kitaev}%
  \BibitemOpen
  \bibfield  {author} {\bibinfo {author} {\bibfnamefont {A.}~\bibnamefont
  {Kitaev}},\ }\bibfield  {title} {\bibinfo {title} {Anyons in an exactly
  solved model and beyond},\ }\href
  {https://doi.org/https://doi.org/10.1016/j.aop.2005.10.005} {\bibfield
  {journal} {\bibinfo  {journal} {Ann. Phys. (NY)}\ }\textbf {\bibinfo {volume}
  {321}},\ \bibinfo {pages} {2} (\bibinfo {year} {2006})},\ \bibinfo {note}
  {january Special Issue}\BibitemShut {NoStop}%
\bibitem [{\citenamefont {Do}\ \emph {et~al.}(2017)\citenamefont {Do},
  \citenamefont {Park}, \citenamefont {Yoshitake}, \citenamefont {Nasu},
  \citenamefont {Motome}, \citenamefont {Kwon}, \citenamefont {Adroja},
  \citenamefont {Voneshen}, \citenamefont {Kim}, \citenamefont {Jang},
  \citenamefont {Park}, \citenamefont {Choi},\ and\ \citenamefont
  {Ji}}]{MajoranaKitaevExp}%
  \BibitemOpen
  \bibfield  {author} {\bibinfo {author} {\bibfnamefont {S.-H.}\ \bibnamefont
  {Do}}, \bibinfo {author} {\bibfnamefont {S.-Y.}\ \bibnamefont {Park}},
  \bibinfo {author} {\bibfnamefont {J.}~\bibnamefont {Yoshitake}}, \bibinfo
  {author} {\bibfnamefont {J.}~\bibnamefont {Nasu}}, \bibinfo {author}
  {\bibfnamefont {Y.}~\bibnamefont {Motome}}, \bibinfo {author} {\bibfnamefont
  {Y.~S.}\ \bibnamefont {Kwon}}, \bibinfo {author} {\bibfnamefont {D.~T.}\
  \bibnamefont {Adroja}}, \bibinfo {author} {\bibfnamefont {D.~J.}\
  \bibnamefont {Voneshen}}, \bibinfo {author} {\bibfnamefont {K.}~\bibnamefont
  {Kim}}, \bibinfo {author} {\bibfnamefont {T.~H.}\ \bibnamefont {Jang}},
  \bibinfo {author} {\bibfnamefont {J.~H.}\ \bibnamefont {Park}}, \bibinfo
  {author} {\bibfnamefont {K.-Y.}\ \bibnamefont {Choi}},\ and\ \bibinfo
  {author} {\bibfnamefont {S.}~\bibnamefont {Ji}},\ }\bibfield  {title}
  {\bibinfo {title} {Majorana fermions in the {Kitaev} quantum spin system
  $\alpha$-{RuCl3}},\ }\href {https://doi.org/10.1038/nphys4264} {\bibfield
  {journal} {\bibinfo  {journal} {Nat. Phys.}\ }\textbf {\bibinfo {volume}
  {13}},\ \bibinfo {pages} {1079} (\bibinfo {year} {2017})}\BibitemShut
  {NoStop}%
\bibitem [{\citenamefont {Motome}\ and\ \citenamefont
  {Nasu}(2020)}]{reviewMajorana}%
  \BibitemOpen
  \bibfield  {author} {\bibinfo {author} {\bibfnamefont {Y.}~\bibnamefont
  {Motome}}\ and\ \bibinfo {author} {\bibfnamefont {J.}~\bibnamefont {Nasu}},\
  }\bibfield  {title} {\bibinfo {title} {Hunting {Majorana} fermions in
  {Kitaev} magnets},\ }\href {https://doi.org/10.7566/JPSJ.89.012002}
  {\bibfield  {journal} {\bibinfo  {journal} {J. Phys. Soc. Japan}\ }\textbf
  {\bibinfo {volume} {89}},\ \bibinfo {pages} {012002} (\bibinfo {year}
  {2020})}\BibitemShut {NoStop}%
\bibitem [{\citenamefont {Nayak}\ \emph {et~al.}(2008)\citenamefont {Nayak},
  \citenamefont {Simon}, \citenamefont {Stern}, \citenamefont {Freedman},\ and\
  \citenamefont {Das~Sarma}}]{Nayak08}%
  \BibitemOpen
  \bibfield  {author} {\bibinfo {author} {\bibfnamefont {C.}~\bibnamefont
  {Nayak}}, \bibinfo {author} {\bibfnamefont {S.~H.}\ \bibnamefont {Simon}},
  \bibinfo {author} {\bibfnamefont {A.}~\bibnamefont {Stern}}, \bibinfo
  {author} {\bibfnamefont {M.}~\bibnamefont {Freedman}},\ and\ \bibinfo
  {author} {\bibfnamefont {S.}~\bibnamefont {Das~Sarma}},\ }\bibfield  {title}
  {\bibinfo {title} {Non-abelian anyons and topological quantum computation},\
  }\href {https://doi.org/10.1103/RevModPhys.80.1083} {\bibfield  {journal}
  {\bibinfo  {journal} {Rev. Mod. Phys.}\ }\textbf {\bibinfo {volume} {80}},\
  \bibinfo {pages} {1083} (\bibinfo {year} {2008})}\BibitemShut {NoStop}%
\bibitem [{\citenamefont {Takagi}\ \emph {et~al.}(2019)\citenamefont {Takagi},
  \citenamefont {Takayama}, \citenamefont {Jackeli}, \citenamefont
  {Khaliullin},\ and\ \citenamefont {Nagler}}]{Takagi2019}%
  \BibitemOpen
  \bibfield  {author} {\bibinfo {author} {\bibfnamefont {H.}~\bibnamefont
  {Takagi}}, \bibinfo {author} {\bibfnamefont {T.}~\bibnamefont {Takayama}},
  \bibinfo {author} {\bibfnamefont {G.}~\bibnamefont {Jackeli}}, \bibinfo
  {author} {\bibfnamefont {G.}~\bibnamefont {Khaliullin}},\ and\ \bibinfo
  {author} {\bibfnamefont {S.~E.}\ \bibnamefont {Nagler}},\ }\bibfield  {title}
  {\bibinfo {title} {Concept and realization of {Kitaev} quantum spin
  liquids},\ }\href {https://doi.org/10.1038/s42254-019-0038-2} {\bibfield
  {journal} {\bibinfo  {journal} {Nat. Rev. Phys.}\ }\textbf {\bibinfo {volume}
  {1}},\ \bibinfo {pages} {264} (\bibinfo {year} {2019})}\BibitemShut {NoStop}%
\bibitem [{\citenamefont {Guinea}\ \emph {et~al.}(2010)\citenamefont {Guinea},
  \citenamefont {Katsnelson},\ and\ \citenamefont {Geim}}]{guinea}%
  \BibitemOpen
  \bibfield  {author} {\bibinfo {author} {\bibfnamefont {F.}~\bibnamefont
  {Guinea}}, \bibinfo {author} {\bibfnamefont {M.~I.}\ \bibnamefont
  {Katsnelson}},\ and\ \bibinfo {author} {\bibfnamefont {A.~K.}\ \bibnamefont
  {Geim}},\ }\bibfield  {title} {\bibinfo {title} {Energy gaps and a zero-field
  quantum {Hall} effect in graphene by strain engineering},\ }\href
  {https://doi.org/10.1038/nphys1420} {\bibfield  {journal} {\bibinfo
  {journal} {Nat. Phys.}\ }\textbf {\bibinfo {volume} {6}},\ \bibinfo {pages}
  {30} (\bibinfo {year} {2010})}\BibitemShut {NoStop}%
\bibitem [{\citenamefont {Levy}\ \emph {et~al.}(2010)\citenamefont {Levy},
  \citenamefont {Burke}, \citenamefont {Meaker}, \citenamefont {Panlasigui},
  \citenamefont {Zettl}, \citenamefont {Guinea}, \citenamefont {Neto},\ and\
  \citenamefont {Crommie}}]{exp1}%
  \BibitemOpen
  \bibfield  {author} {\bibinfo {author} {\bibfnamefont {N.}~\bibnamefont
  {Levy}}, \bibinfo {author} {\bibfnamefont {S.~A.}\ \bibnamefont {Burke}},
  \bibinfo {author} {\bibfnamefont {K.~L.}\ \bibnamefont {Meaker}}, \bibinfo
  {author} {\bibfnamefont {M.}~\bibnamefont {Panlasigui}}, \bibinfo {author}
  {\bibfnamefont {A.}~\bibnamefont {Zettl}}, \bibinfo {author} {\bibfnamefont
  {F.}~\bibnamefont {Guinea}}, \bibinfo {author} {\bibfnamefont {A.~H.~C.}\
  \bibnamefont {Neto}},\ and\ \bibinfo {author} {\bibfnamefont {M.~F.}\
  \bibnamefont {Crommie}},\ }\bibfield  {title} {\bibinfo {title}
  {Strain-induced pseudo magnetic fields greater than 300 {Tesla} in graphene
  nanobubbles},\ }\href {https://doi.org/10.1126/science.1191700} {\bibfield
  {journal} {\bibinfo  {journal} {Science}\ }\textbf {\bibinfo {volume}
  {329}},\ \bibinfo {pages} {544} (\bibinfo {year} {2010})}\BibitemShut
  {NoStop}%
\bibitem [{\citenamefont {Yeh}\ \emph {et~al.}(2011)\citenamefont {Yeh},
  \citenamefont {Teague}, \citenamefont {Yeom}, \citenamefont {Standley},
  \citenamefont {Wu}, \citenamefont {Boyd},\ and\ \citenamefont
  {Bockrath}}]{exp2}%
  \BibitemOpen
  \bibfield  {author} {\bibinfo {author} {\bibfnamefont {N.-C.}\ \bibnamefont
  {Yeh}}, \bibinfo {author} {\bibfnamefont {M.-L.}\ \bibnamefont {Teague}},
  \bibinfo {author} {\bibfnamefont {S.}~\bibnamefont {Yeom}}, \bibinfo {author}
  {\bibfnamefont {B.}~\bibnamefont {Standley}}, \bibinfo {author}
  {\bibfnamefont {R.-P.}\ \bibnamefont {Wu}}, \bibinfo {author} {\bibfnamefont
  {D.}~\bibnamefont {Boyd}},\ and\ \bibinfo {author} {\bibfnamefont
  {M.}~\bibnamefont {Bockrath}},\ }\bibfield  {title} {\bibinfo {title}
  {Strain-induced pseudo-magnetic fields and charging effects on {CVD}-grown
  graphene},\ }\href
  {https://doi.org/https://doi.org/10.1016/j.susc.2011.03.025} {\bibfield
  {journal} {\bibinfo  {journal} {Surf. Sci.}\ }\textbf {\bibinfo {volume}
  {605}},\ \bibinfo {pages} {1649} (\bibinfo {year} {2011})}\BibitemShut
  {NoStop}%
\bibitem [{\citenamefont {Lu}\ \emph {et~al.}(2012)\citenamefont {Lu},
  \citenamefont {Neto},\ and\ \citenamefont {Loh}}]{exp3}%
  \BibitemOpen
  \bibfield  {author} {\bibinfo {author} {\bibfnamefont {J.}~\bibnamefont
  {Lu}}, \bibinfo {author} {\bibfnamefont {A.~H.~C.}\ \bibnamefont {Neto}},\
  and\ \bibinfo {author} {\bibfnamefont {K.~P.}\ \bibnamefont {Loh}},\
  }\bibfield  {title} {\bibinfo {title} {Transforming moir{\'e} blisters into
  geometric graphene nano-bubbles},\ }\href
  {https://doi.org/10.1038/ncomms1818} {\bibfield  {journal} {\bibinfo
  {journal} {Nat. Commun.}\ }\textbf {\bibinfo {volume} {3}},\ \bibinfo {pages}
  {823} (\bibinfo {year} {2012})}\BibitemShut {NoStop}%
\bibitem [{\citenamefont {Li}\ \emph {et~al.}(2015)\citenamefont {Li},
  \citenamefont {Bai}, \citenamefont {Yin}, \citenamefont {Qiao}, \citenamefont
  {Wang},\ and\ \citenamefont {He}}]{exp4}%
  \BibitemOpen
  \bibfield  {author} {\bibinfo {author} {\bibfnamefont {S.-Y.}\ \bibnamefont
  {Li}}, \bibinfo {author} {\bibfnamefont {K.-K.}\ \bibnamefont {Bai}},
  \bibinfo {author} {\bibfnamefont {L.-J.}\ \bibnamefont {Yin}}, \bibinfo
  {author} {\bibfnamefont {J.-B.}\ \bibnamefont {Qiao}}, \bibinfo {author}
  {\bibfnamefont {W.-X.}\ \bibnamefont {Wang}},\ and\ \bibinfo {author}
  {\bibfnamefont {L.}~\bibnamefont {He}},\ }\bibfield  {title} {\bibinfo
  {title} {Observation of unconventional splitting of {Landau} levels in
  strained graphene},\ }\href {https://doi.org/10.1103/PhysRevB.92.245302}
  {\bibfield  {journal} {\bibinfo  {journal} {Phys. Rev. B}\ }\textbf {\bibinfo
  {volume} {92}},\ \bibinfo {pages} {245302} (\bibinfo {year}
  {2015})}\BibitemShut {NoStop}%
\bibitem [{\citenamefont {Gomes}\ \emph {et~al.}(2012)\citenamefont {Gomes},
  \citenamefont {Mar}, \citenamefont {Ko}, \citenamefont {Guinea},\ and\
  \citenamefont {Manoharan}}]{exp8}%
  \BibitemOpen
  \bibfield  {author} {\bibinfo {author} {\bibfnamefont {K.~K.}\ \bibnamefont
  {Gomes}}, \bibinfo {author} {\bibfnamefont {W.}~\bibnamefont {Mar}}, \bibinfo
  {author} {\bibfnamefont {W.}~\bibnamefont {Ko}}, \bibinfo {author}
  {\bibfnamefont {F.}~\bibnamefont {Guinea}},\ and\ \bibinfo {author}
  {\bibfnamefont {H.~C.}\ \bibnamefont {Manoharan}},\ }\bibfield  {title}
  {\bibinfo {title} {Designer {Dirac} fermions and topological phases in
  molecular graphene},\ }\href {https://doi.org/10.1038/nature10941} {\bibfield
   {journal} {\bibinfo  {journal} {Nature}\ }\textbf {\bibinfo {volume}
  {483}},\ \bibinfo {pages} {306} (\bibinfo {year} {2012})}\BibitemShut
  {NoStop}%
\bibitem [{\citenamefont {Rechtsman}\ \emph {et~al.}(2013)\citenamefont
  {Rechtsman}, \citenamefont {Zeuner}, \citenamefont {T{\"u}nnermann},
  \citenamefont {Nolte}, \citenamefont {Segev},\ and\ \citenamefont
  {Szameit}}]{Rechtsman2013}%
  \BibitemOpen
  \bibfield  {author} {\bibinfo {author} {\bibfnamefont {M.~C.}\ \bibnamefont
  {Rechtsman}}, \bibinfo {author} {\bibfnamefont {J.~M.}\ \bibnamefont
  {Zeuner}}, \bibinfo {author} {\bibfnamefont {A.}~\bibnamefont
  {T{\"u}nnermann}}, \bibinfo {author} {\bibfnamefont {S.}~\bibnamefont
  {Nolte}}, \bibinfo {author} {\bibfnamefont {M.}~\bibnamefont {Segev}},\ and\
  \bibinfo {author} {\bibfnamefont {A.}~\bibnamefont {Szameit}},\ }\bibfield
  {title} {\bibinfo {title} {Strain-induced pseudomagnetic field and photonic
  {Landau} levels in dielectric structures},\ }\href
  {https://doi.org/10.1038/nphoton.2012.302} {\bibfield  {journal} {\bibinfo
  {journal} {Nat. Photon.}\ }\textbf {\bibinfo {volume} {7}},\ \bibinfo {pages}
  {153} (\bibinfo {year} {2013})}\BibitemShut {NoStop}%
\bibitem [{\citenamefont {Wen}\ \emph {et~al.}(2019)\citenamefont {Wen},
  \citenamefont {Qiu}, \citenamefont {Qi}, \citenamefont {Ye}, \citenamefont
  {Ke}, \citenamefont {Zhang},\ and\ \citenamefont {Liu}}]{exp7}%
  \BibitemOpen
  \bibfield  {author} {\bibinfo {author} {\bibfnamefont {X.}~\bibnamefont
  {Wen}}, \bibinfo {author} {\bibfnamefont {C.}~\bibnamefont {Qiu}}, \bibinfo
  {author} {\bibfnamefont {Y.}~\bibnamefont {Qi}}, \bibinfo {author}
  {\bibfnamefont {L.}~\bibnamefont {Ye}}, \bibinfo {author} {\bibfnamefont
  {M.}~\bibnamefont {Ke}}, \bibinfo {author} {\bibfnamefont {F.}~\bibnamefont
  {Zhang}},\ and\ \bibinfo {author} {\bibfnamefont {Z.}~\bibnamefont {Liu}},\
  }\bibfield  {title} {\bibinfo {title} {Acoustic {Landau} quantization and
  quantum-{Hall}-like edge states},\ }\href
  {https://doi.org/10.1038/s41567-019-0446-3} {\bibfield  {journal} {\bibinfo
  {journal} {Nat. Phys.}\ }\textbf {\bibinfo {volume} {15}},\ \bibinfo {pages}
  {352} (\bibinfo {year} {2019})}\BibitemShut {NoStop}%
\bibitem [{\citenamefont {Bellec}\ \emph {et~al.}(2020)\citenamefont {Bellec},
  \citenamefont {Poli}, \citenamefont {Kuhl}, \citenamefont {Mortessagne},\
  and\ \citenamefont {Schomerus}}]{exp5}%
  \BibitemOpen
  \bibfield  {author} {\bibinfo {author} {\bibfnamefont {M.}~\bibnamefont
  {Bellec}}, \bibinfo {author} {\bibfnamefont {C.}~\bibnamefont {Poli}},
  \bibinfo {author} {\bibfnamefont {U.}~\bibnamefont {Kuhl}}, \bibinfo {author}
  {\bibfnamefont {F.}~\bibnamefont {Mortessagne}},\ and\ \bibinfo {author}
  {\bibfnamefont {H.}~\bibnamefont {Schomerus}},\ }\bibfield  {title} {\bibinfo
  {title} {Observation of supersymmetric pseudo-{Landau} levels in strained
  microwave graphene},\ }\href {https://doi.org/10.1038/s41377-020-00351-2}
  {\bibfield  {journal} {\bibinfo  {journal} {Light Sci. Appl.}\ }\textbf
  {\bibinfo {volume} {9}},\ \bibinfo {pages} {146} (\bibinfo {year}
  {2020})}\BibitemShut {NoStop}%
\bibitem [{\citenamefont {Jamadi}\ \emph {et~al.}(2020)\citenamefont {Jamadi},
  \citenamefont {Rozas}, \citenamefont {Salerno}, \citenamefont
  {Mili{\'{c}}evi{\'{c}}}, \citenamefont {Ozawa}, \citenamefont {Sagnes},
  \citenamefont {Lema{\^i}tre}, \citenamefont {Le~Gratiet}, \citenamefont
  {Harouri}, \citenamefont {Carusotto}, \citenamefont {Bloch},\ and\
  \citenamefont {Amo}}]{exp6}%
  \BibitemOpen
  \bibfield  {author} {\bibinfo {author} {\bibfnamefont {O.}~\bibnamefont
  {Jamadi}}, \bibinfo {author} {\bibfnamefont {E.}~\bibnamefont {Rozas}},
  \bibinfo {author} {\bibfnamefont {G.}~\bibnamefont {Salerno}}, \bibinfo
  {author} {\bibfnamefont {M.}~\bibnamefont {Mili{\'{c}}evi{\'{c}}}}, \bibinfo
  {author} {\bibfnamefont {T.}~\bibnamefont {Ozawa}}, \bibinfo {author}
  {\bibfnamefont {I.}~\bibnamefont {Sagnes}}, \bibinfo {author} {\bibfnamefont
  {A.}~\bibnamefont {Lema{\^i}tre}}, \bibinfo {author} {\bibfnamefont
  {L.}~\bibnamefont {Le~Gratiet}}, \bibinfo {author} {\bibfnamefont
  {A.}~\bibnamefont {Harouri}}, \bibinfo {author} {\bibfnamefont
  {I.}~\bibnamefont {Carusotto}}, \bibinfo {author} {\bibfnamefont
  {J.}~\bibnamefont {Bloch}},\ and\ \bibinfo {author} {\bibfnamefont
  {A.}~\bibnamefont {Amo}},\ }\bibfield  {title} {\bibinfo {title} {Direct
  observation of photonic {Landau} levels and helical edge states in strained
  honeycomb lattices},\ }\href {https://doi.org/10.1038/s41377-020-00377-6}
  {\bibfield  {journal} {\bibinfo  {journal} {Light Sci. Appl.}\ }\textbf
  {\bibinfo {volume} {9}},\ \bibinfo {pages} {144} (\bibinfo {year}
  {2020})}\BibitemShut {NoStop}%
\bibitem [{\citenamefont {Poli}\ \emph {et~al.}(2014)\citenamefont {Poli},
  \citenamefont {Arkinstall},\ and\ \citenamefont {Schomerus}}]{cas2014}%
  \BibitemOpen
  \bibfield  {author} {\bibinfo {author} {\bibfnamefont {C.}~\bibnamefont
  {Poli}}, \bibinfo {author} {\bibfnamefont {J.}~\bibnamefont {Arkinstall}},\
  and\ \bibinfo {author} {\bibfnamefont {H.}~\bibnamefont {Schomerus}},\
  }\bibfield  {title} {\bibinfo {title} {Degeneracy doubling and sublattice
  polarization in strain-induced pseudo-{Landau} levels},\ }\href
  {https://doi.org/10.1103/PhysRevB.90.155418} {\bibfield  {journal} {\bibinfo
  {journal} {Phys. Rev. B}\ }\textbf {\bibinfo {volume} {90}},\ \bibinfo
  {pages} {155418} (\bibinfo {year} {2014})}\BibitemShut {NoStop}%
\bibitem [{\citenamefont {Rachel}\ \emph {et~al.}(2016)\citenamefont {Rachel},
  \citenamefont {G\"othel}, \citenamefont {Arovas},\ and\ \citenamefont
  {Vojta}}]{rachel2016}%
  \BibitemOpen
  \bibfield  {author} {\bibinfo {author} {\bibfnamefont {S.}~\bibnamefont
  {Rachel}}, \bibinfo {author} {\bibfnamefont {I.}~\bibnamefont {G\"othel}},
  \bibinfo {author} {\bibfnamefont {D.~P.}\ \bibnamefont {Arovas}},\ and\
  \bibinfo {author} {\bibfnamefont {M.}~\bibnamefont {Vojta}},\ }\bibfield
  {title} {\bibinfo {title} {Strain-induced {Landau} levels in arbitrary
  dimensions with an exact spectrum},\ }\href
  {https://doi.org/10.1103/PhysRevLett.117.266801} {\bibfield  {journal}
  {\bibinfo  {journal} {Phys. Rev. Lett.}\ }\textbf {\bibinfo {volume} {117}},\
  \bibinfo {pages} {266801} (\bibinfo {year} {2016})}\BibitemShut {NoStop}%
\bibitem [{\citenamefont {Pnueli}(1994)}]{pnuelli1994}%
  \BibitemOpen
  \bibfield  {author} {\bibinfo {author} {\bibfnamefont {A.}~\bibnamefont
  {Pnueli}},\ }\bibfield  {title} {\bibinfo {title} {Spinors and scalars on
  {Riemann} surfaces},\ }\href {https://doi.org/10.1088/0305-4470/27/4/028}
  {\bibfield  {journal} {\bibinfo  {journal} {J. Phys. A}\ }\textbf {\bibinfo
  {volume} {27}},\ \bibinfo {pages} {1345} (\bibinfo {year}
  {1994})}\BibitemShut {NoStop}%
\bibitem [{\citenamefont {Wen}\ and\ \citenamefont {Zee}(1992)}]{WZ}%
  \BibitemOpen
  \bibfield  {author} {\bibinfo {author} {\bibfnamefont {X.~G.}\ \bibnamefont
  {Wen}}\ and\ \bibinfo {author} {\bibfnamefont {A.}~\bibnamefont {Zee}},\
  }\bibfield  {title} {\bibinfo {title} {Shift and spin vector: New topological
  quantum numbers for the {Hall} fluids},\ }\href
  {https://doi.org/10.1103/PhysRevLett.69.953} {\bibfield  {journal} {\bibinfo
  {journal} {Phys. Rev. Lett.}\ }\textbf {\bibinfo {volume} {69}},\ \bibinfo
  {pages} {953} (\bibinfo {year} {1992})}\BibitemShut {NoStop}%
\bibitem [{\citenamefont {Nguyen}\ \emph {et~al.}(2018)\citenamefont {Nguyen},
  \citenamefont {Golkar}, \citenamefont {Roberts},\ and\ \citenamefont
  {Son}}]{Nguyen2018}%
  \BibitemOpen
  \bibfield  {author} {\bibinfo {author} {\bibfnamefont {D.~X.}\ \bibnamefont
  {Nguyen}}, \bibinfo {author} {\bibfnamefont {S.}~\bibnamefont {Golkar}},
  \bibinfo {author} {\bibfnamefont {M.~M.}\ \bibnamefont {Roberts}},\ and\
  \bibinfo {author} {\bibfnamefont {D.~T.}\ \bibnamefont {Son}},\ }\bibfield
  {title} {\bibinfo {title} {Particle-hole symmetry and composite fermions in
  fractional quantum {Hall} states},\ }\href
  {https://doi.org/10.1103/PhysRevB.97.195314} {\bibfield  {journal} {\bibinfo
  {journal} {Phys. Rev. B}\ }\textbf {\bibinfo {volume} {97}},\ \bibinfo
  {pages} {195314} (\bibinfo {year} {2018})}\BibitemShut {NoStop}%
\bibitem [{\citenamefont {Golkar}\ \emph {et~al.}(2014)\citenamefont {Golkar},
  \citenamefont {Roberts},\ and\ \citenamefont {Son}}]{Golkar2014}%
  \BibitemOpen
  \bibfield  {author} {\bibinfo {author} {\bibfnamefont {S.}~\bibnamefont
  {Golkar}}, \bibinfo {author} {\bibfnamefont {M.~M.}\ \bibnamefont
  {Roberts}},\ and\ \bibinfo {author} {\bibfnamefont {D.~T.}\ \bibnamefont
  {Son}},\ }\bibfield  {title} {\bibinfo {title} {Effective field theory of
  relativistic quantum hall systems},\ }\href
  {https://doi.org/10.1007/JHEP12(2014)138} {\bibfield  {journal} {\bibinfo
  {journal} {J. High Energy Phys.}\ }\textbf {\bibinfo {volume} {2014}}\bibinfo
   {number} { (12)},\ \bibinfo {pages} {138}}\BibitemShut {NoStop}%
\bibitem [{\citenamefont {Castro~Neto}\ \emph {et~al.}(2009)\citenamefont
  {Castro~Neto}, \citenamefont {Guinea}, \citenamefont {Peres}, \citenamefont
  {Novoselov},\ and\ \citenamefont {Geim}}]{graphenereview}%
  \BibitemOpen
\bibfield  {number} {  }\bibfield  {author} {\bibinfo {author} {\bibfnamefont
  {A.~H.}\ \bibnamefont {Castro~Neto}}, \bibinfo {author} {\bibfnamefont
  {F.}~\bibnamefont {Guinea}}, \bibinfo {author} {\bibfnamefont {N.~M.~R.}\
  \bibnamefont {Peres}}, \bibinfo {author} {\bibfnamefont {K.~S.}\ \bibnamefont
  {Novoselov}},\ and\ \bibinfo {author} {\bibfnamefont {A.~K.}\ \bibnamefont
  {Geim}},\ }\bibfield  {title} {\bibinfo {title} {The electronic properties of
  graphene},\ }\href {https://doi.org/10.1103/RevModPhys.81.109} {\bibfield
  {journal} {\bibinfo  {journal} {Rev. Mod. Phys.}\ }\textbf {\bibinfo {volume}
  {81}},\ \bibinfo {pages} {109} (\bibinfo {year} {2009})}\BibitemShut
  {NoStop}%
\bibitem [{\citenamefont {Iorio}\ and\ \citenamefont {Pais}(2015)}]{strain1}%
  \BibitemOpen
  \bibfield  {author} {\bibinfo {author} {\bibfnamefont {A.}~\bibnamefont
  {Iorio}}\ and\ \bibinfo {author} {\bibfnamefont {P.}~\bibnamefont {Pais}},\
  }\bibfield  {title} {\bibinfo {title} {Revisiting the gauge fields of
  strained graphene},\ }\href {https://doi.org/10.1103/PhysRevD.92.125005}
  {\bibfield  {journal} {\bibinfo  {journal} {Phys. Rev. D}\ }\textbf {\bibinfo
  {volume} {92}},\ \bibinfo {pages} {125005} (\bibinfo {year}
  {2015})}\BibitemShut {NoStop}%
\bibitem [{\citenamefont {de~Juan}\ \emph {et~al.}(2012)\citenamefont
  {de~Juan}, \citenamefont {Sturla},\ and\ \citenamefont
  {Vozmediano}}]{juan2012}%
  \BibitemOpen
  \bibfield  {author} {\bibinfo {author} {\bibfnamefont {F.}~\bibnamefont
  {de~Juan}}, \bibinfo {author} {\bibfnamefont {M.}~\bibnamefont {Sturla}},\
  and\ \bibinfo {author} {\bibfnamefont {M.~A.~H.}\ \bibnamefont
  {Vozmediano}},\ }\bibfield  {title} {\bibinfo {title} {Space dependent fermi
  velocity in strained graphene},\ }\href
  {https://doi.org/10.1103/PhysRevLett.108.227205} {\bibfield  {journal}
  {\bibinfo  {journal} {Phys. Rev. Lett.}\ }\textbf {\bibinfo {volume} {108}},\
  \bibinfo {pages} {227205} (\bibinfo {year} {2012})}\BibitemShut {NoStop}%
\bibitem [{\citenamefont {de~Juan}\ \emph {et~al.}(2013)\citenamefont
  {de~Juan}, \citenamefont {Ma\~nes},\ and\ \citenamefont
  {Vozmediano}}]{juan2013}%
  \BibitemOpen
  \bibfield  {author} {\bibinfo {author} {\bibfnamefont {F.}~\bibnamefont
  {de~Juan}}, \bibinfo {author} {\bibfnamefont {J.~L.}\ \bibnamefont
  {Ma\~nes}},\ and\ \bibinfo {author} {\bibfnamefont {M.~A.~H.}\ \bibnamefont
  {Vozmediano}},\ }\bibfield  {title} {\bibinfo {title} {Gauge fields from
  strain in graphene},\ }\href {https://doi.org/10.1103/PhysRevB.87.165131}
  {\bibfield  {journal} {\bibinfo  {journal} {Phys. Rev. B}\ }\textbf {\bibinfo
  {volume} {87}},\ \bibinfo {pages} {165131} (\bibinfo {year}
  {2013})}\BibitemShut {NoStop}%
\bibitem [{\citenamefont {Zubkov}\ and\ \citenamefont
  {Volovik}(2015)}]{zubkov2015}%
  \BibitemOpen
  \bibfield  {author} {\bibinfo {author} {\bibfnamefont {M.~A.}\ \bibnamefont
  {Zubkov}}\ and\ \bibinfo {author} {\bibfnamefont {G.~E.}\ \bibnamefont
  {Volovik}},\ }\bibfield  {title} {\bibinfo {title} {Emergent gravity in
  graphene},\ }\href {https://doi.org/10.1088/1742-6596/607/1/012020}
  {\bibfield  {journal} {\bibinfo  {journal} {J. Phys. Conf. Ser}\ }\textbf
  {\bibinfo {volume} {607}},\ \bibinfo {pages} {012020} (\bibinfo {year}
  {2015})}\BibitemShut {NoStop}%
\bibitem [{\citenamefont {Schomerus}\ and\ \citenamefont
  {Halpern}(2013)}]{Henning2013}%
  \BibitemOpen
  \bibfield  {author} {\bibinfo {author} {\bibfnamefont {H.}~\bibnamefont
  {Schomerus}}\ and\ \bibinfo {author} {\bibfnamefont {N.~Y.}\ \bibnamefont
  {Halpern}},\ }\bibfield  {title} {\bibinfo {title} {Parity anomaly and
  {Landau}-level lasing in strained photonic honeycomb lattices},\ }\href
  {https://doi.org/10.1103/PhysRevLett.110.013903} {\bibfield  {journal}
  {\bibinfo  {journal} {Phys. Rev. Lett.}\ }\textbf {\bibinfo {volume} {110}},\
  \bibinfo {pages} {013903} (\bibinfo {year} {2013})}\BibitemShut {NoStop}%
\bibitem [{\citenamefont {Wagner}\ \emph {et~al.}(2022)\citenamefont {Wagner},
  \citenamefont {de~Juan},\ and\ \citenamefont {Nguyen}}]{wagner2022}%
  \BibitemOpen
  \bibfield  {author} {\bibinfo {author} {\bibfnamefont {G.}~\bibnamefont
  {Wagner}}, \bibinfo {author} {\bibfnamefont {F.}~\bibnamefont {de~Juan}},\
  and\ \bibinfo {author} {\bibfnamefont {D.~X.}\ \bibnamefont {Nguyen}},\
  }\bibfield  {title} {\bibinfo {title} {{Landau levels in curved space
  realized in strained graphene}},\ }\href
  {https://doi.org/10.21468/SciPostPhysCore.5.2.029} {\bibfield  {journal}
  {\bibinfo  {journal} {SciPost Phys. Core}\ }\textbf {\bibinfo {volume} {5}},\
  \bibinfo {pages} {029} (\bibinfo {year} {2022})}\BibitemShut {NoStop}%
\bibitem [{\citenamefont {Hofstadter}(1976)}]{Hof1976}%
  \BibitemOpen
  \bibfield  {author} {\bibinfo {author} {\bibfnamefont {D.~R.}\ \bibnamefont
  {Hofstadter}},\ }\bibfield  {title} {\bibinfo {title} {Energy levels and wave
  functions of {Bloch} electrons in rational and irrational magnetic fields},\
  }\href {https://doi.org/10.1103/PhysRevB.14.2239} {\bibfield  {journal}
  {\bibinfo  {journal} {Phys. Rev. B}\ }\textbf {\bibinfo {volume} {14}},\
  \bibinfo {pages} {2239} (\bibinfo {year} {1976})}\BibitemShut {NoStop}%
\bibitem [{\citenamefont {Gonz\'alez}\ \emph {et~al.}(1992)\citenamefont
  {Gonz\'alez}, \citenamefont {Guinea},\ and\ \citenamefont
  {Vozmediano}}]{GGV92}%
  \BibitemOpen
  \bibfield  {author} {\bibinfo {author} {\bibfnamefont {J.}~\bibnamefont
  {Gonz\'alez}}, \bibinfo {author} {\bibfnamefont {F.}~\bibnamefont {Guinea}},\
  and\ \bibinfo {author} {\bibfnamefont {M.~A.~H.}\ \bibnamefont
  {Vozmediano}},\ }\bibfield  {title} {\bibinfo {title} {Continuum
  approximation to fullerene molecules},\ }\href
  {https://doi.org/10.1103/PhysRevLett.69.172} {\bibfield  {journal} {\bibinfo
  {journal} {Phys. Rev. Lett.}\ }\textbf {\bibinfo {volume} {69}},\ \bibinfo
  {pages} {172} (\bibinfo {year} {1992})}\BibitemShut {NoStop}%
\bibitem [{\citenamefont {Singh}\ \emph {et~al.}(2012)\citenamefont {Singh},
  \citenamefont {Manni}, \citenamefont {Reuther}, \citenamefont {Berlijn},
  \citenamefont {Thomale}, \citenamefont {Ku}, \citenamefont {Trebst},\ and\
  \citenamefont {Gegenwart}}]{Singh2012}%
  \BibitemOpen
  \bibfield  {author} {\bibinfo {author} {\bibfnamefont {Y.}~\bibnamefont
  {Singh}}, \bibinfo {author} {\bibfnamefont {S.}~\bibnamefont {Manni}},
  \bibinfo {author} {\bibfnamefont {J.}~\bibnamefont {Reuther}}, \bibinfo
  {author} {\bibfnamefont {T.}~\bibnamefont {Berlijn}}, \bibinfo {author}
  {\bibfnamefont {R.}~\bibnamefont {Thomale}}, \bibinfo {author} {\bibfnamefont
  {W.}~\bibnamefont {Ku}}, \bibinfo {author} {\bibfnamefont {S.}~\bibnamefont
  {Trebst}},\ and\ \bibinfo {author} {\bibfnamefont {P.}~\bibnamefont
  {Gegenwart}},\ }\bibfield  {title} {\bibinfo {title} {Relevance of the
  heisenberg-kitaev model for the honeycomb lattice iridates
  ${A}_{2}{\mathrm{iro}}_{3}$},\ }\href
  {https://doi.org/10.1103/PhysRevLett.108.127203} {\bibfield  {journal}
  {\bibinfo  {journal} {Phys. Rev. Lett.}\ }\textbf {\bibinfo {volume} {108}},\
  \bibinfo {pages} {127203} (\bibinfo {year} {2012})}\BibitemShut {NoStop}%
\bibitem [{\citenamefont {Lahtinen}\ \emph {et~al.}(2012)\citenamefont
  {Lahtinen}, \citenamefont {Ludwig}, \citenamefont {Pachos},\ and\
  \citenamefont {Trebst}}]{vortexKitaev1}%
  \BibitemOpen
  \bibfield  {author} {\bibinfo {author} {\bibfnamefont {V.}~\bibnamefont
  {Lahtinen}}, \bibinfo {author} {\bibfnamefont {A.~W.~W.}\ \bibnamefont
  {Ludwig}}, \bibinfo {author} {\bibfnamefont {J.~K.}\ \bibnamefont {Pachos}},\
  and\ \bibinfo {author} {\bibfnamefont {S.}~\bibnamefont {Trebst}},\
  }\bibfield  {title} {\bibinfo {title} {Topological liquid nucleation induced
  by vortex-vortex interactions in {Kitaev}'s honeycomb model},\ }\href
  {https://doi.org/10.1103/PhysRevB.86.075115} {\bibfield  {journal} {\bibinfo
  {journal} {Phys. Rev. B}\ }\textbf {\bibinfo {volume} {86}},\ \bibinfo
  {pages} {075115} (\bibinfo {year} {2012})}\BibitemShut {NoStop}%
\bibitem [{\citenamefont {Lahtinen}\ \emph {et~al.}(2008)\citenamefont
  {Lahtinen}, \citenamefont {Kells}, \citenamefont {Carollo}, \citenamefont
  {Stitt}, \citenamefont {Vala},\ and\ \citenamefont {Pachos}}]{vortexKitaev2}%
  \BibitemOpen
  \bibfield  {author} {\bibinfo {author} {\bibfnamefont {V.}~\bibnamefont
  {Lahtinen}}, \bibinfo {author} {\bibfnamefont {G.}~\bibnamefont {Kells}},
  \bibinfo {author} {\bibfnamefont {A.}~\bibnamefont {Carollo}}, \bibinfo
  {author} {\bibfnamefont {T.}~\bibnamefont {Stitt}}, \bibinfo {author}
  {\bibfnamefont {J.}~\bibnamefont {Vala}},\ and\ \bibinfo {author}
  {\bibfnamefont {J.~K.}\ \bibnamefont {Pachos}},\ }\bibfield  {title}
  {\bibinfo {title} {Spectrum of the non-abelian phase in {Kitaev}'s honeycomb
  lattice model},\ }\href
  {https://doi.org/https://doi.org/10.1016/j.aop.2007.12.009} {\bibfield
  {journal} {\bibinfo  {journal} {Ann. Phys. (NY)}\ }\textbf {\bibinfo {volume}
  {323}},\ \bibinfo {pages} {2286} (\bibinfo {year} {2008})}\BibitemShut
  {NoStop}%
\bibitem [{\citenamefont {Kells}\ \emph {et~al.}(2008)\citenamefont {Kells},
  \citenamefont {Bolukbasi}, \citenamefont {Lahtinen}, \citenamefont
  {Slingerland}, \citenamefont {Pachos},\ and\ \citenamefont
  {Vala}}]{vortexKitaev3}%
  \BibitemOpen
  \bibfield  {author} {\bibinfo {author} {\bibfnamefont {G.}~\bibnamefont
  {Kells}}, \bibinfo {author} {\bibfnamefont {A.~T.}\ \bibnamefont
  {Bolukbasi}}, \bibinfo {author} {\bibfnamefont {V.}~\bibnamefont {Lahtinen}},
  \bibinfo {author} {\bibfnamefont {J.~K.}\ \bibnamefont {Slingerland}},
  \bibinfo {author} {\bibfnamefont {J.~K.}\ \bibnamefont {Pachos}},\ and\
  \bibinfo {author} {\bibfnamefont {J.}~\bibnamefont {Vala}},\ }\bibfield
  {title} {\bibinfo {title} {Topological degeneracy and vortex manipulation in
  {Kitaev}'s honeycomb model},\ }\href
  {https://doi.org/10.1103/PhysRevLett.101.240404} {\bibfield  {journal}
  {\bibinfo  {journal} {Phys. Rev. Lett.}\ }\textbf {\bibinfo {volume} {101}},\
  \bibinfo {pages} {240404} (\bibinfo {year} {2008})}\BibitemShut {NoStop}%
\bibitem [{\citenamefont {Chen}\ and\ \citenamefont
  {Nussinov}(2008)}]{Chen:2008}%
  \BibitemOpen
  \bibfield  {author} {\bibinfo {author} {\bibfnamefont {H.-D.}\ \bibnamefont
  {Chen}}\ and\ \bibinfo {author} {\bibfnamefont {Z.}~\bibnamefont
  {Nussinov}},\ }\bibfield  {title} {\bibinfo {title} {Exact results of the
  {Kitaev} model on a hexagonal lattice: spin states, string and brane
  correlators, and anyonic excitations},\ }\href
  {https://doi.org/10.1088/1751-8113/41/7/075001} {\bibfield  {journal}
  {\bibinfo  {journal} {J. Phys. A}\ }\textbf {\bibinfo {volume} {41}},\
  \bibinfo {pages} {075001} (\bibinfo {year} {2008})}\BibitemShut {NoStop}%
\bibitem [{\citenamefont {Keil}\ \emph {et~al.}(2016)\citenamefont {Keil},
  \citenamefont {Poli}, \citenamefont {Heinrich}, \citenamefont {Arkinstall},
  \citenamefont {Weihs}, \citenamefont {Schomerus},\ and\ \citenamefont
  {Szameit}}]{Keil16}%
  \BibitemOpen
  \bibfield  {author} {\bibinfo {author} {\bibfnamefont {R.}~\bibnamefont
  {Keil}}, \bibinfo {author} {\bibfnamefont {C.}~\bibnamefont {Poli}}, \bibinfo
  {author} {\bibfnamefont {M.}~\bibnamefont {Heinrich}}, \bibinfo {author}
  {\bibfnamefont {J.}~\bibnamefont {Arkinstall}}, \bibinfo {author}
  {\bibfnamefont {G.}~\bibnamefont {Weihs}}, \bibinfo {author} {\bibfnamefont
  {H.}~\bibnamefont {Schomerus}},\ and\ \bibinfo {author} {\bibfnamefont
  {A.}~\bibnamefont {Szameit}},\ }\bibfield  {title} {\bibinfo {title}
  {Universal sign control of coupling in tight-binding lattices},\ }\href
  {https://doi.org/10.1103/PhysRevLett.116.213901} {\bibfield  {journal}
  {\bibinfo  {journal} {Phys. Rev. Lett.}\ }\textbf {\bibinfo {volume} {116}},\
  \bibinfo {pages} {213901} (\bibinfo {year} {2016})}\BibitemShut {NoStop}%
\bibitem [{\citenamefont {Gorbar}\ and\ \citenamefont {Gusynin}(2008)}]{GG08}%
  \BibitemOpen
  \bibfield  {author} {\bibinfo {author} {\bibfnamefont {E.}~\bibnamefont
  {Gorbar}}\ and\ \bibinfo {author} {\bibfnamefont {V.}~\bibnamefont
  {Gusynin}},\ }\bibfield  {title} {\bibinfo {title} {Gap generation for
  {Dirac} fermions on {Lobachevsky} plane in a magnetic field},\ }\href
  {https://doi.org/10.1016/j.aop.2007.11.005} {\bibfield  {journal} {\bibinfo
  {journal} {Ann. Phys. (NY)}\ }\textbf {\bibinfo {volume} {323}},\ \bibinfo
  {pages} {2132 } (\bibinfo {year} {2008})}\BibitemShut {NoStop}%
\end{thebibliography}%

\end{document}